\documentclass[fleqn]{article}
\usepackage{fleqn}
\usepackage{graphicx}
\usepackage{epsfig}
\usepackage{amsmath}
\usepackage{amssymb}
\usepackage{calc}
\usepackage{color}
\newcommand{\beq}{\begin{equation}}
\newcommand{\eeq}{\end{equation}}
\newcommand{\bea}{\begin{eqnarray}}
\newcommand{\eea}{\end{eqnarray}}

\usepackage{tikz}
\usetikzlibrary{shapes,snakes}
\usepackage{amsmath,amssymb}
\usepackage{verbatim}

\newcommand{\du}{\node[rectangle] {$\downarrow\uparrow$}}
\newcommand{\de}{\node[rectangle] {$\downarrow$\textcolor{white}{$\uparrow$}}}
\newcommand{\ue}{\node[rectangle] {$\uparrow$\textcolor{white}{$\uparrow$}}}
\newcommand{\ee}{\node[rectangle] {\textcolor{white}{$\uparrow\downarrow$}}}

\usepackage{hyperref}
\hypersetup{
 pdfauthor={P\'eter L\'evay and Fr\'ederic Holweck},
 pdftitle={Embedding qubits into fermionic Fock space},
 pdfproducer={LaTeX with hyperref},
 pdffitwindow=false,
 pdfstartview={FitH},
 colorlinks=true,
 linkcolor={blue},
 citecolor={blue},
 filecolor={blue},
 urlcolor={blue}
}
\begin{document}
\Large
\begin{center}
{\bf  Embedding qubits into fermionic Fock space, peculiarities of the four-qubit case}
\end{center}
\large
\vspace*{-.1cm}
\begin{center}
 P\'eter L\'evay$^{1}$ and Fr\'ederic Holweck$^{2}$
\end{center}
\vspace*{-.4cm} \normalsize
\begin{center}

$^{1}$Department of Theoretical Physics, Institute of Physics, Budapest University of\\
Technology and Economics and MTA-BME Condensed Matter Research Group, H-1521 Budapest, Hungary

$^{2}$Laboratoire IRTES-M3M,
Universit\'e de Technologie de Belfort-Montb\'eliard, 90010 Belfort Cedex, France

\vspace*{.0cm}

\vspace*{.2cm} (16 February 2015)

\end{center}

\vspace*{-.3cm} \noindent \hrulefill

\vspace*{.1cm} \noindent {\bf Abstract:} We give a fermionic Fock
space description of embedded entangled qubits. Within this
framework the problem of classification of pure state entanglement
boils down to the problem of classifying spinors. The usual notion
of separable states turns out to be just a special case of the one
of pure spinors. By using the notion of single, double and mixed
occupancy representation with intertwiners relating them a natural
physical interpretation of embedded qubits is found. As an
application of these ideas one can make a physical sound meaning
of {\it some} of the direct sum structures showing up in the
context of the so called Black-Hole/Qubit Correspondence. We
discuss how the usual invariants for qubits serving as measures of
entanglement can be obtained from invariants for spinors in an
elegant manner. In particular a detailed case study for recovering
the invariants for four-qubits within a spinorial framework is
presented. We also observe that reality conditions on complex
spinors defining Majorana spinors for embedded qubits boil down to
self conjugate states under the Wootters spin flip operation.
Finally we conduct a study on the explicit structure of
$Spin(16,\mathbb{C})$ invariant polynomials related to the
structure of possible measures of entanglement for fermionic
systems with $8$ modes. Here we find an algebraically independent
generating set of the generalized SLOCC invariants and calculate
their restriction to the dense orbit. We point out the special
role the largest exceptional group $E_8$ is playing in these
considerations.

 \vspace*{.3cm} \noindent
{\bf PACS:} 02.40.Dr, 03.65.Ud, 03.65.Ta \\
{\bf Keywords:}  Quantum entanglement, representation theory,
invariants.\\ \hspace*{1.95cm} --

\vspace*{-.2cm} \noindent \hrulefill

\section{Introduction}

It is well-known that a system of $n$ distinguishable qubits can naturally be embedded into a  system
of $n$ fermions with $2n$ modes. This idea has widely been used with applications in quantum chemistry\cite{Borland,Ruskai}, in studies concerning the relationship between spin systems and fermionic ones\cite{Cirac}, the QMA-completenes of the N-representability problem\cite{Liu} , entanglement classification and canonical forms\cite{LevVran,VranLev,Chen,Chen2}, ground state properties of fermionic systems\cite{Ocko}
and
the so-called Black-Hole/Qubit Correspondence (BHQC)\cite{BHQC}.

In a recent study
it has been shown\cite{SarLev1} that in order to gain
further insight into the structure of such entangled systems
it is rewarding to regard them as embedded ones into the {\it  full} fermionic Fock space.
Physically this means that apart from the usual protocols of preserving the number of fermions we should also allow ones for manipulating such systems via changing the fermion number.
This idea leads us to the notion of generalized Bogoliubov transformations\cite{SarLev1}.
Though the physical significance of this idea is yet to be explored even at this stage it
makes it possible
to regard the classification problem of entanglement types under stochastic local operations and classical communication (SLOCC)\cite{Dur} as a special case of a problem well-known to mathematicians as the problem of classifying {\it spinors}\cite{Cartan,Chevalley,Trautman,Trautman2,Igusa}.
This observation makes a step towards establishing a unified framework for understanding  entanglement
properties of quantum systems consisiting of subsystems with both distinguishable and indistinguishable
constituents.
The aim of the present paper is to take a further step in this direction and present a systematic study of $n$-qubit systems as ones living inside the {\it full} fermionic Fock space with
particular emphasis put on four-qubit systems.

The motivation for embarking in this investigation is twofold.
The first is to provide further clues for understanding the structure of pure state multipartite entanglement measures as the ones arising from invariants for spinors.
These are homogeneous polynomials in the complex amplitudes of the pure fermionic states,
capable of identifying certain types of entanglement in Fock space.
It has been observed\cite{SarLev1} that
when considering in the broader fermionic context
some of the multiqubit polynomial invariants
have a more transparent geometric and algebraic structure than
in the original multiqubit one.
Making use of the full Fock space these structures  are easy to identify
and straightforward to do calculations with.
With our investigations we would like to shed some further light on such issues by working out explicitely the embedded four-qubit case.
Our treatise can also be regarded as an elaboration on some of the ideas presented in Ref.\cite{Chen} in connection with four qubit states regarded as fermionic ones.

The second source of motivation is coming from the BHQC\cite{BHQC}.
In this context it was observed\cite{DuffFerrara,Levayfano} that it is useful to reinterpret some of the irreducible representation spaces
of groups like $E_7({\mathbb C})$ and $SO(12,\mathbb{C})$ and many others\cite{Octonions}
(whose real forms are showing up in black hole entropy formulas of certain supergravity theories)
as ones composed of a certain number of qubits.
Since the invariants associated to these representation spaces have the  physical meaning as the black hole entropy
this qubit picture lends itself naturally to an interpretation of black hole entropy
as a manifestation of some sort of entanglement.
In simple special cases this interpretation has turned out to be a useful one for obtaining further insight
into the structure of black hole solutions in supergravity\cite{BHQC}. However, for the more complicated cases no conventional entanglement based reinterpretation have been found.

In order to illustrate this problem arising in BHQC let us consider the so called R-R subsector of $N=8$ supergravity\cite{BHQC}. In this case we have $32$ charges describing the winding configurations of certain extended objects like membranes and strings on noncontractible cycles of extra dimesions. These charges are transforming  according to the spinor representation of the group $SO(6,6)$ a real form of the group
$SO(12,\mathbb{C})$. Then the representation space is $\mathcal{H}={\mathbb R}^{32}$.   Naively one would think that since $32=2^5$ this representation space is amenable
to a $5$ real qubit (or rebit) interpretation, i.e $\mathcal{H}=V_A\otimes V_{B}\otimes V_C\otimes V_{D}\otimes V_E$
where $V_{A,B,C,D,E}$ are five copies of ${\mathbb R}^2$s. Of course this interpretation is wrong since it cannot accomodate the $SO(6,6)$ action.
However, one can accomodate this action via employing $6$ real qubits\footnote{Clearly one can permute the labels $A,B,C,D,E,F$ for convenience provided we leave the incidence structure of Eq. (\ref{pelda}) intact.
This incidence structure is that of a terahedron. For alternative labelling of these qubits see Refs.\cite{DuffFerrara,Levayfano}.
} to build up the spinor representation  space $\mathcal{H}$
in the form\cite{DuffFerrara,Levayfano}
\beq
\mathcal{H}=V_{ACF}\oplus V_{ADE}\oplus V_{BCE}\oplus V_{BDF},\qquad V_{ACF}\equiv V_A\otimes V_C\otimes V_F\qquad {\rm e.t.c.}
\label{pelda}
\eeq
\noindent
However, the problem with this structure is that at first sight it is not amenable to any conventional
quantum information theoretic interpretation as an entangled system. The reason for this is simple: the presence of the direct sums.
In the language of representation theory this problem can be rephrased as the decomposition of $SO(6,6)$ under the subgroup $SL(2)_A\times SL(2)_B\times SL(2)_C\times SL(2)_D\times SL(2)_E\times SL(2)_F$ namely
\beq
32\to (2,1,2,1,1,2)\oplus(2,1,1,2,2,1)\oplus (1,2,2,1,2,1)\oplus (1,2,1,2,1,2).
\label{pelda2}
\eeq
\noindent
In this language the problem is that unlike the doublets (qubits) in the conventional theory of quantum entanglement we cannot make sense of the singlets.
In Appendix C. of the review paper of Borsten et.al.\cite{Octonions} many more examples of that kind have been discussed. Some of them are related to representation spaces of exceptional groups, and having direct relevance to string theory and supergravity.
According to Ref.\cite{Octonions} the unusual "tripartite entanglement of six qubits" of Eq. (\ref{pelda})
can be given the conventional quantum information theoretic interpretation by regarding it as a subspace
of six entangled {\it qutrits}. The weak point of this suggestion is that there is no physically sound reason
why we should restrict our attention to this particular $32$ dimensional  subspace inside the $3^6$
dimensional space rather then to any other one. For this proposal to make sense one should somehow specify the physical protocols which are represented by those transformations of this qutrit space
that leave this $32$ dimensional subspace invariant.
Similar criticism should be applied to the remaining systems of Appendix C. of Ref.\cite{Octonions}.
In this paper we show that at least in the special case of systems amenable to a fermionic Fock space description structures like the one of Eq. (\ref{pelda}) can be made a natural interpretation as embedded qubit systems which avoids the problem posed above.

The plan of this paper is as follows.
In Section 2. we summarize the basic material on fermionic Fock space and indentify the generalized SLOCC group as
$\mathbb{C}^{\times}\times {\rm
Spin}(2N,\mathbb{C})$. Here ${\rm Spin}(2N,\mathbb{C})$ is accomodating generalized Bogoliubov transformations. In Section 3. we reconsider some of the ideas of Ref.\cite{SarLev1} on entanglement in fermionic Fock space. We show how ordinary SLOCC transformations\cite{Dur} are accomodated within the formalism and how the SLOCC classification problem is recovered within the framework of a more general mathematical problem, namely the classification of spinors. Here we also comment on the special role of {\it pure} spinors giving a natural generalization of the notion of separable states.
In Section 4. we begin our investigation of embedded $n$-qubit systems
by studying the single occupancy representation. We observe that the largest subgroup of the fermionic SLOCC group $GL(2n,\mathbb{C})$ leaving invariant the single occupancy subspace is the group
$\tilde{G}=S_n\ltimes G$ where $G=GL(2,\mathbb{C})^{\times n}$ and $S_n$ is the symmetric group.
 We will make use of this in our analysis of four-qubit states when searching for permutation invariant combinations of the usual SLOCC invariants within a spinorial formalism.
In Section 5. we embark in an elementary discussion on different ways for embedding qubits.
We go through in detail the basic structures we come accross up to four qubits.
Here we introduce intertwiners relating the double, single and mixed occupancy representations occurring in these descriptions of qubits. An interesting observation on the role of these intertwiners as maps related to the mirror map of string theory relating the IIA and IIB duality frames is presented. These investigations make it clear how one should make sense {\it some} of the direct sum structures (the occurrence of singlets) in the BHQC.
Section 6. is devoted to a detailed study of invariants and covariants paying special attention to the spinorial case
yielding embedded four-qubit states.
For the four qubit case the structure of the basic spinorial covariant is of block diagonal form
the blocks related to the two-partite reduced density matrices via a conjugation operation.
This is the spinorial generalization\cite{SarLevlegujabb}
of the well-known Wootters spin flip operation\cite{Wootters}.
We point out that the notion of Wootters self conjugate spinor is just  the notion of a {\it Majorana} spinor.
In closing this section we recover the basic four-qubit $\tilde{G}$ invariants in a spinorial framework.
In Section 7. we conduct a study on the explicit structure of $Spin(16,\mathbb{C})$ invariant polynomials
i.e. on the structure of generalized SLOCC invariants for fermionic systems with $8$ modes.
We point out the special role the largest exceptional group $E_8$ is playing in this respect.
We introduce a fermionic state which is a representative of the semisimple orbit, depending on $8$ complex parameters. Then in terms of these parameters for this particular state we calculate the values of $8$
polynomial invariants, which form a basis of generators of the ring of ${\rm Spin}(16,\mathbb{C})$ invariant polynomials.
We show that the resulting polynomials in $8$ variables can neatly be expressed in terms of $8$ algebraically independent polynomials which are invariant under the Weyl group of $E_8$. 
Our conclusion and comments are left for Section. 8.

\section{Fermionic Fock Space}

In this section we summarize results concerning spinors in a fermionic Fock space language\cite{SarLev1,Cartan, Chevalley,Trautman,Trautman2}.
Let $V$ be an $N$ dimensional complex vector space and $V^{\ast}$ its dual.
We regard $V\simeq {\mathbb {C}}^N$ with $\{e_i\}, i=1,2,\dots N$ the canonical basis and $\{e^i\}$ is the dual basis. Elements of $V$ will be called {\it one particle states}.
We tacitly assume that $V$ is a finite dimensional Hilbert space also equipped with a Hermitian inner product, but at first we will not make use of this extra structure untill Sec. 6.6.
We also introduce the $2N$ dimensional vector space
\beq
\mathcal {V}\equiv V\oplus V^{\ast}
\label{nu}
\eeq
\noindent
with
basis $\{e^{I}\}\equiv\{e_i,e^j\},\quad I=1,\dots N,N+1,\dots 2N$.
An element of $\mathcal{V}$ is of the form $x=v+\alpha$ where $v$ is a vector
and $\alpha$ is a linear form with $v=v^ie_i$ and $\alpha=\alpha_je^j$.
According to the method of second quantization to any element $x\equiv x_Ie^I\in\mathcal{V}$ one can associate a linear operator $\hat{x}$ acting on a $2^N$ dimensional complex vector space $\mathcal {F}$ called the {\it fermionic Fock space} ${\mathcal F}$ as follows.
Take the exterior (Grassmann) algebra $\wedge^{\bullet}V^{\ast}$
where
\beq
\wedge^{\bullet}V^{\ast}=\mathbb{C}\oplus V^{\ast}\oplus \wedge^2V^{\ast}\oplus\cdots\oplus \wedge^NV^{\ast}.
\label{grass}
\eeq
\noindent
Then the Fock space is defined as
\beq
\mathcal{F}\equiv \wedge^{\bullet}V^{\ast}\otimes (\wedge^N V)^{-1/2}.
\label{fock}
\eeq
\noindent
The origin of the last factor will be explained later (see Eq.(\ref{fertraf})).
Temporarily the reader should regard ${\mathcal F}$ merely as the Grassmann algebra (\ref{grass}) based on $V^{\ast}$.
Now the operator $\hat{x}=x_I\hat{e}^I=v^i\hat{e}_i+\alpha_j\hat{e}^j$ acting on $\mathcal{F}$ is obtained by the assignment

\beq e^i\mapsto \hat{e}^i\equiv e^i\wedge,\qquad e_i\mapsto
\hat{e}_j\equiv {\iota}_{e_i} \label{repext} \eeq \noindent i.e.
the basis vectors are mapped to the operators of exterior and
interior multiplication. Defining $\{\hat{x},\hat{y}\}\equiv
\hat{x}\hat{y}+\hat{y}\hat{x}$ we have \beq
\{\hat{e}^i,\hat{e}_j\}={\delta^{i}}_j\hat{1},\quad
\{\hat{e}^i,\hat{e}^j\}= \{\hat{e}_i,\hat{e}_j\}=0
\label{fermicomm1} \eeq \noindent which are the usual fermionic
{\it anticommutation} relations.

The one-dimensional subspace  $\wedge^0V^{\ast}=\mathbb{C}$
corresponds to the ray of the {\it vacuum state} denoted by $\vert
0\rangle$. The operators $\hat{e}^i$ and $\hat{e}_j$ are the {\it
creation} and {\it annihilation} operators. For later convenience
we redefine these  as \beq \hat{e}^i\equiv \hat{p}^i,\qquad
\hat{e}_j\equiv \hat{n}_j \label{redef} \eeq \noindent with \beq
\{\hat{p}^i,\hat{n}_j\}={\delta^i}_j\hat{1},\quad
\{\hat{p}^i,\hat{p}^j\}= \{\hat{n}_i,\hat{n}_j\}=0.
\label{ujraanti} \eeq \noindent The algebra above will be called
the CAR algebra (the algebra of canonical anticommutation
relations). This algebra can compactly be expressed as \beq
\{\hat{e}_I,\hat{e}_J\}=g_{IJ}\hat{1},\qquad
g_{IJ}=\begin{pmatrix}0&I\\I&0\end{pmatrix}\label{kompaktalak}
\eeq \noindent where $g_{IJ}$ is a $2N\times 2N$ matrix with
$N\times N$ blocks and $I$ is the $N\times N$ identity matrix.

Clearly \beq \hat{n}_j\vert 0\rangle =0 \label{vacuumprop} \eeq
\noindent encapsulates the defining property of the vacuum, namely
that it contains no "particles" or "excitations" at all. On the
other hand the state \beq \hat{p}^i\vert 0\rangle \label{egy} \eeq
\noindent represents a single "particle" which is  in the $i$th
"mode". Similarly states of the form \beq \hat{p}^i\hat{p}^j\vert
0\rangle,\qquad \hat{p}^i\hat{p}^j\hat{p}^k\vert
0\rangle,\qquad\dots,\qquad \hat{p}^1\hat{p}^2\cdots
\hat{p}^N\vert 0\rangle,\qquad i<j<k\qquad {\rm e.t.c.}
\label{particles}\eeq \noindent are the
two, three  $\dots N$ "particle" states. Generally the  ${N\choose
k}$-dimensional $k$-particle subspace is spanned by the basis
vectors $\hat{p}^{i_1}\hat{p}^{i_2}\cdots \hat{p}^{i_k}\vert
0\rangle$ with $1\leq i_1<i_2\cdots <i_k\leq N$. It then follows
that an arbitrary state of $\mathcal{F}$ can be written in the
form \beq \vert \psi\rangle\equiv \hat{\Psi}\vert 0\rangle,\qquad
\hat{\Psi}\equiv\sum_{k=0}^N\sum_{i_1i_2\cdots
i_k=1}^N\frac{1}{k!}\psi^{(k)}_{i_1i_2\dots
i_k}\hat{p}^{i_1}\hat{p}^{i_2}\cdots \hat{p}^{i_k}. \label{steexp}
\eeq \noindent Here the $k$th order totally antisymmetric tensors
$\psi^{(k)}_{i_1i_2\cdots i_k}$ encapsulate the complex amplitudes
of the $k$-particle subspace. An element $\vert
\psi\rangle\in\mathcal{F}$ is called a {\it spinor}.

One can alternatively define the linear combinations \beq
\hat{\Gamma}_{i}=\hat{p}^i+\hat{n}_i,\qquad
\hat{\Gamma}_{i+N}=\hat{p}^i-\hat{n}_i,\qquad i=1,2,\dots ,N
\label{gammam} \eeq \noindent satisfying \beq
\{\hat{\Gamma}_I,\hat{\Gamma}_J\}=2\eta_{IJ}\hat{1}, \qquad
\eta_{IJ}
=\begin{pmatrix}\delta_{ij}&0\\0&-\delta_{ij}\end{pmatrix} ,\qquad
I,J=1,\dots ,2N,\quad i,j=1,\dots ,N. \label{diraccomm} \eeq
\noindent The matrix representatives of the $\hat{\Gamma}_I$
operators correspond to the usual gamma matrices in physics. Let
us now define the operator \beq \hat{\Gamma}\equiv
[\hat{n}_1,\hat{p}^1]
[\hat{n}_2,\hat{p}^2][\hat{n}_3,\hat{p}^3]\cdots
[\hat{n}_N,\hat{p}^N]=(-1)^{N(N-1)/2}\hat{\Gamma}_1\hat{\Gamma}_2\hat{\Gamma}_3\cdots\hat{\Gamma}_N.
\label{gammaot} \eeq \noindent It is easy to check that
$\hat{\Gamma}^2=\hat{1}$ hence the eigenvalues of this operator
are $\pm 1$. Spinors $\vert\psi_{\pm}\rangle$ which are
eigenvectors of $\hat{\Gamma}$ corresponding to the eigenvalues
$\pm 1$ are called {\it Weyl spinors} of positive and negative
{\it helicity} or {\it chirality}. One chan check that in
Eq.(\ref{steexp}) spinors of positive chirality have terms with an
{\it even}, and negative chirality have terms with an {\it odd}
number of creation operators. Hence we have the decomposition \beq
\mathcal{F}=\mathcal{F}_+\oplus\mathcal{F}_-.\label{fockdecomp}
\eeq \noindent

Let us now take the operators
$\hat{x}=x^I\hat{e}_I=\alpha_i\hat{p}^i+v^j\hat{n}_j$ and
$\hat{y}=y^J\hat{e}_J=\beta_i\hat{p}^i+w^j\hat{n}_j$ answering the
corresponding vectors $x$ and $y$ having the same expansions with
hats removed. Then \beq
\{\hat{x},\hat{y}\}=g(x,y)\hat{1}=g_{IJ}x^Iy^J\hat{1}=(\alpha_iw^i+v^j\beta_j)\hat{1}.
\label{biliforma} \eeq \noindent Here
$g:\mathcal{V}\times\mathcal{V}\to \mathbb{C}$ is a nondegenerate
symmetric bilinear form with matrix $g_{IJ}$ known from
Eq.(\ref{kompaktalak}). The group of transformations which leave
this form invariant is the orthogonal group
$O(\mathcal{V},g)\equiv O(2N,\mathbb{C})$. We take its connected
component to the identity which is $SO(2N,\mathbb{C})$. We have
\beq g(\mathcal{S}(x),\mathcal{S}(y))=g(x,y),\qquad \mathcal{S}\in
SO(2N,\mathbb{C}). \label{invrot} \eeq \noindent Using matrices
this equation yields \beq
\mathcal{S}^tg\mathcal{S}=g.\label{matrica} \eeq \noindent Writing
$\mathcal{S}=e^s$ where $s\in\mathfrak{s}\mathfrak{o}(2N)$ and using
Eq.(\ref{invrot}) by taking the infinitesimal version of
Eq.(\ref{matrica}) one can see that $s$ can be parametrized as
\beq s=\begin{pmatrix}A&B\\C&-A^{t}\end{pmatrix},\qquad
C^t=-C,\quad B^t=-B \label{para} \eeq \noindent where $A,B,C$ are
$N\times N$ matrices with $B^t$ refers to the transposed matrix of
$B$. One can also regard an $\mathcal{S}$ as a transformation
acting on the operators $\hat{x}$. Then combining
Eqs.(\ref{biliforma}) and (\ref{invrot}) one can see that elements
of $SO(2N,\mathbb{C})$ also leave the CAR algebra invariant. These
transformations will be called generalized Bogoliubov
transformations.

We would also like to have an action of these generalized
Bogoliubov transformations on our Fock space $\mathcal{F}$. The
usual way to define this action is via introducing operators
$\hat{S}$ that are mapped to the transformations $\mathcal{S}$ via
the relation \beq
\hat{S}\hat{x}\hat{S}^{-1}=\mathcal{S}(\hat{x}),\qquad
x\in\mathcal{V}, \quad \mathcal{S}\in SO(2N,\mathbb{C}).
\label{spindefi} \eeq \noindent Here
$\mathcal{S}(\hat{x})=x^I\mathcal{S}(\hat{e}_I)=x^I\hat{e}_J{\mathcal{S}^J}_I$
where the matrix ${\mathcal{S}^J}_I$ is the exponential of the
matrix given by Eq.(\ref{para}). It is well-known that the set of
such transformations $\hat{S}$ gives the double cover of
$SO(2N,\mathbb{C})$ which is the group ${\rm
Spin}(2N,\mathbb{C})$. Let us write the infinitesimal version of
Eq.(\ref{spindefi}) in the form \beq
[\hat{s},\hat{x}]=s(\hat{x}),\qquad \hat{s}\in {\rm
spin}(2N),\quad s\in {\rm spin}(2N)\simeq\mathfrak{s}\mathfrak{o}(2N). \label{infispin} \eeq \noindent
Now a calculation shows that \beq
\hat{s}=\frac{1}{2}{A_i}^j[\hat{p}^i,\hat{n}_j]+\frac{1}{2}B_{ij}\hat{p}^i\hat{p}^j+\frac{1}{2}C^{ij}\hat{n}_i\hat{n}_j.
\label{sexplicit} \eeq \noindent Clearly for the action
$\hat{S}=e^{\hat{s}}$ the subspaces $\mathcal{F}_{\pm}$ are
invariant ones. In the following transformations of the form \beq
\vert\psi_{\pm}\rangle\mapsto \lambda
e^{\hat{s}}\vert\psi_{\pm}\rangle,\qquad
(\lambda,e^{\hat{s}})\in\mathbb{C}^{\times}\times {\rm
Spin}(2N,\mathbb{C}),\quad
\vert\psi_{\pm}\rangle\in\mathcal{F}_{\pm} \label{genslocc} \eeq
\noindent will be called {\it  generalized SLOCC transformations}.
The rationale for also including the group $\mathbb{C}^{\times}$
of nonzero complex numbers will be given in Eq.(\ref{sl2n}).
Notice that the set of generalized SLOCC transformations is
respecting the chirality of Weyl spinors.

\bigskip
\section{Entanglement}
\bigskip
In this section we summarize results on entanglement in fermionic Fock space\cite{SarLev1}.

The main advantage of the subspaces $\mathcal{F}_{\pm}$ is that we
can embed into them the state spaces of a large variety of
multipartite entangled systems taken together with the action of
their respective SLOCC groups. In order to see this one just have
to realize that the particle number conserving subgroup of the
generalized SLOCC group is obtained by setting $B=C=0$ in
Eq.(\ref{sexplicit}). Now we have \beq
\hat{s}=\frac{1}{2}{A_i}^j[\hat{p}^i,\hat{n}_j]={A_i}^j\hat{p}^i\hat{n}_j
-\frac{1}{2}{\rm Tr}(A)\hat{1}.\label{sloccgr} \eeq \noindent The
exponential of this is \beq \hat{S}=e^{\hat{s}}=e^{-{\rm
Tr}A/2}e^{{A_i}^j\hat{p}^i\hat{n}_j}. \label{valodislocc} \eeq
\noindent Let us now consider the action of $\hat{S}$ on a
$k$-particle subspace \beq \hat{S}\vert{\psi}^{(k)}\rangle=
\frac{1}{k!}{\psi^{(k)}}_{i_1i_2\dots
i_k}(\hat{S}\hat{p}^{i_1}\hat{S}^{-1})
(\hat{S}\hat{p}^{i_2}\hat{S}^{-1})\cdots
(\hat{S}\hat{p}^{i_k}\hat{S}^{-1})\hat{S}\vert 0\rangle \eeq
\noindent According to Eq.(\ref{spindefi}) we have \beq
\hat{S}\hat{p}^i\hat{S}^{-1}=\hat{p}^j{\mathcal{S}_j}^i,\qquad
{\mathcal{S}}=e^{{A}}\in GL(N,\mathbb{C}) \eeq \noindent hence
\beq \hat{S}\vert{\psi}^{(k)}\rangle=\frac{1}{k!}\psi^{\prime
(k)}_{j_1\dots j_k} \hat{p}^{j_1}\cdots \hat{p}^{j_k}\vert
0\rangle,\quad \psi^{\prime (k)}_{j_1\dots j_k}
=({\rm{Det}\mathcal{S}})^{-1/2}{\mathcal{S}_{j_1}}^{i_1}\cdots
{\mathcal{S}_{j_k}}^{i_k} {\psi^{(k)}}_{i_1\dots i_k}.
\label{fertraf} \eeq \noindent Here we have used the identity
$e^{{\rm Tr}A}={\rm Det}\mathcal{S}$ where $\mathcal{S}=e^A$.
Eq.(\ref{fertraf}) shows that apart from the extra term $({\rm
Det}\mathcal{S})^{-1/2}$ the totally antisymmetric tensor
$\psi^{(k)}_{i_1i_2\cdots i_k}$ incorporating the ${N\choose k}$
complex amplitudes transforms via $N$ identical copies of the
usual fermionic SLOCC group i.e. $GL(N,\mathbb{C})$ well-known
from the theory of fermionic entanglement. Notice also that the
presence of the extra term clearly shows that the fermionic Fock
space should be the one of Eq.(\ref{fock}) we started our
considerations with.

Let us fix a spinor $\vert\psi\rangle\in{\mathcal{F}}$ and define
its {\it annihilator subspace} $\mathcal{M}_{\psi}$ of
$\mathcal{V}$ as the set of vectors $x\in\mathcal{V}$ such that
their corresponding operators $\hat{x}$ annihilate
$\vert\psi\rangle$. \beq
\mathcal{M}_{\psi}\equiv\{x\in\mathcal{V}\vert\quad\hat{x}\vert\psi\rangle
=0\}. \label{annihsubs} \eeq \noindent From Eq.(\ref{biliforma})
it follows that if $x,y\in\mathcal{M}_{\psi}$ then $g(x,y)=0$.
Hence $\mathcal{M}_{\psi}$ is a {\it totally isotropic subspace}
of $\mathcal{V}$. Clearly due to the structure of our bilinear
form $g$ the maximal dimension of a totally isotropic subspace is
$N$. A spinor $\vert\psi\rangle$ such that $\mathcal{M}_{\psi}$ is
a {\it maximal} totally isotropic subspace of $\mathcal{V}$ is
called a {\it pure spinor}. (Cartan calls them {\it simple
spinors}, Chevalley calls them {\it pure spinors}. Here we follow
the conventions based on the English literature and will call them
pure spinors.)

First of all notice that all of the spinors showing up in the
sequence of Eqs.(\ref{vacuumprop})-(\ref{particles}) are pure.
Indeed, take for instance
$\vert\psi\rangle=\hat{p}^{k+1}\hat{p}^{k+2}\cdots\hat{p}^N\vert
0\rangle$ where $k=0,1,\dots N$. Then \beq \mathcal{M}_{\psi}={\rm
span}\{{n}_1,{n}_2,\dots {n}_k,{p}^{k+1},{p}^{k+2},\dots {p}^N\}.
\label{peldaann} \eeq \noindent Since any pair of operators
corresponding to vectors taken from this set is pairwise
anticommuting, according to Eq.(\ref{biliforma}) this is a totally
isotropic subspace, with maximal dimension $N$. From this it
follows that all the pure spinors of the form \beq
\hat{p}^{i_1}\hat{p}^{i_2}\cdots \hat{p}^{i_k}\vert
0\rangle\leftrightarrow e^{i_1}\wedge e^{i_2}\wedge\cdots\wedge
e^{i_k}\label{slaterdet} \eeq \noindent are {\it Slater
determinants}. Since in the usual theory of fermionic entanglement
(with fixed particle number) the states corresponding to the rays
of Slater determinants are called {\it separable} we conclude that
in the realm of generalized SLOCC transformations the separable
states should be identified with the pure spinors. It is important
to note that {\it pure spinors are Weyl} and there is a one-to-one
correspondence between the rays of pure spinors
and the set of maximally totally isotropic subspaces\cite{Trautman}.
An arbitrary pure spinor
can always be represented in the form \beq \vert\psi^{\rm
pure}\rangle=\lambda e^{\hat{B}} \hat{p}^{i_1}\hat{p}^{i_2}\cdots
\hat{p}^{i_k}\vert 0\rangle,\qquad
\hat{B}=\frac{1}{2}B_{ij}\hat{p}^i\hat{p}^j,\quad B_{ij}=-B_{ji}
\label{puregeneral} \eeq \noindent for some $k=0,1,\dots N$ and
$\lambda\in\mathbb{C}^{\times}$. We will refer to the content of
this equation as the fact that a pure spinor is the so called {\it
B-transform} of a Slater determinant. Spinors that are not pure
will be called {\it entangled}.

Classification of entanglement types in fermionic Fock space
amounts to finding the generalized SLOCC classes, i.e. finding the
orbit structure under the group action of Eq.(\ref{genslocc}).
Since the nontrivial subgroup of this group is ${\rm
Spin}(2N,\mathbb{C})$ and this group respects the chirality of the
spinors one can obtain generalized SLOCC orbits for Weyl spinors
of either type i.e. $\mathcal{F}_+$ or $\mathcal{F}_-$. In the
mathematics literature finding
the generalized SLOCC classes via determining a representative
state from each orbit and its stabilizer is called the {\it
classification problem of spinors}.
It is known that for $N=1,2,3$ every spinor is pure\cite{Igusa,Trautman2}. It means that
the action of the group ${\rm Spin}(2N,\mathbb{C})$ on the space
of Weyl spinors of say positive chirality is transitive. From the
physical point of view it means that there are no entangled states
in the fermionic Fock space for $1,2$ or $3$ single particle
states. The classification problem of spinors was solved by Igusa\cite{Igusa}
for $N=4,5,6$, by Popov\cite{Popov} for $N=7$, and by Antonyan and Elashvili\cite{Antonyan2} for $N=8$ .
These results give the full
orbit structure of entangled states. For $N>8$ coarse
classification schemes have been proposed based on the notion of
the nullity of a spinor\cite{Trautman2}. The {\it nullity} is just the dimension
of the subspace of $\mathcal{V}$ characterized by vectors giving
rise to operators annihilating an entangled state
$\vert\psi\rangle$.

\bigskip
\section{Embedded qubits}
\bigskip

We have seen that the generalized SLOCC group contains naturally
the ordinary fermionic SLOCC group. These groups are acting on
state spaces which are representing quantum systems with {\it
indistinguishable constituents}. One can however, relax this
restriction. Our formalism based on fermionic Fock spaces is also
capable of incorporating systems with {\it distinguishable
constituents}. In this paper we consider the possibility of
incorporating $n$-qubit systems with particular emphasis put on the four-qubit case.

Let $N=2n$, hence our Hilbert space of one-particle states is now
{\it even} dimensional. In this case it is convenient to introduce
a new labelling for the single particle basis states \beq
\{e_1,e_2,\dots e_n,e_{n+1},e_{n+2},\dots e_{2n}\}=
\{e_1,e_2,\dots e_n,e_{\overline{1}},e_{\overline{2}},\dots
e_{\overline{n}}\}. \label{ujracim} \eeq \noindent Let us now
consider $\mathcal{F}^{(n)}$, the ${2n\choose n}$ dimensional
$n$-particle subspace of $\mathcal{F}$. An $n$-fermion state with
$2n$ one-particles states can be written in the form \beq \vert
Z\rangle=\frac{1}{n!}Z_{i_1i_2\dots
i_n}\hat{p}^{i_1}\hat{p}^{i_2}\cdots \hat{p}^{i_n}\vert 0\rangle.
\label{Z} \eeq \noindent Under the SLOCC subgroup of
Eq.(\ref{valodislocc}) the amplitudes of this state transform
as\beq Z_{i_1\dots i_n}\mapsto
({\rm{Det}\mathcal{S}})^{-1/2}{\mathcal{S}_{j_1}}^{i_1}\cdots
{\mathcal{S}_{j_n }}^{i_n} {Z}_{i_1\dots i_n}\equiv
{S_{j_1}}^{i_1}\cdots {S_{j_n }}^{i_n} {Z}_{i_1\dots i_n},
\label{fertrafn} \eeq \noindent where \beq {S_j}^i\equiv ({\rm
Det}\mathcal{S})^{-\frac{1}{2n}}{\mathcal{S}_j}^i\in
SL(2n,\mathbb{C}). \label{sl2n} \eeq \noindent Hence in this
special case the  (\ref{valodislocc}) subgroup of transformations
coming from the group $\hat{S}\in{\rm Spin}(4n,\mathbb{C})$ with
$B_{ij}=C^{ij}=0$ will {\it not} produce the full SLOCC group
$GL(2n,\mathbb{C})$ only an $SL(2n,\mathbb{C})$ subgroup. Luckily,
in Eq.(\ref{genslocc}) we defined the {\it generalized SLOCC
group} as $\mathbb{C}^{\times}\times {\rm Spin}(4n,\mathbb{C})$.
Thanks to this extra $\mathbb{C}^{\times}$ even in this special
case our generalized SLOCC group will contain the ordinary SLOCC
one, namely $GL(2n,\mathbb{C})$. Notice however, that for
$k$-fermion states with $k\neq n$ this subtlety for obtaining the
full SLOCC group is not needed.\footnote{This factor of
$\mathbb{C}^{\times}$ is also needed to be in accord with the
classification of spinors for $N=6$, i.e. $n=3$.  As it is
well-known from the classification theory of prehomogeneous vector
spaces in this case we have a dense orbit of the group
$GL(6,\mathbb{C})\simeq \mathbb{C}^{\times}\times
SL(6,\mathbb{C})$ on the $3$-fermion state space
$\wedge^3\mathbb{C}^6$. This orbit is just the fermionic
generalization of the GHZ orbit known for three-qubits.}

From the set of basis vectors of $\mathcal{F}^{(n)}$ we choose a special subset containing merely $2^n$ elements as follows
\beq
\hat{p}^1\hat{p}^2\cdots\hat{p}^n\vert 0\rangle,\quad
\hat{p}^1\hat{p}^2\cdots\hat{p}^{\overline{n}}\vert 0\rangle,\quad\dots,\quad
\hat{p}^{\overline{1}}\hat{p}^{\overline{2}}\cdots\hat{p}^n\vert 0\rangle,\quad \hat{p}^{\overline{1}}\hat{p}^{\overline{2}}\cdots\hat{p}^{\overline{n}}\vert 0\rangle.
\label{nqubsub}
\eeq
\noindent
These basis vectors will be spanning the state space of embedded $n$-qubit
states. Indeed, let
\beq
\vert\psi\rangle=\sum_{\mu_1,\dots,\mu_n=0,1}\psi_{\mu_1\dots\mu_n}\vert\mu_1\dots\mu_n\rangle\label{nqub}
\eeq
\noindent
be an $n$-qubit state, i.e. an element of $\mathbb{C}^{2^n}$.
Let us now define a map
\beq
f:\mathbb{C}^{2^n}\to \wedge^n\mathbb{C}^{2n}\simeq\mathcal{F}^{(n)}\label{fmap}
\eeq
\noindent
as follows
\beq
\vert\psi\rangle\mapsto \vert Z_{\psi}\rangle=
\left(\psi_{00\dots 0}\hat{p}^1\hat{p}^2\cdots\hat{p}^n+
\psi_{00\dots 1}
\hat{p}^1\hat{p}^2\cdots\hat{p}^{\overline{n}}+\cdots +
\psi_{11\dots 1}
\hat{p}^{\overline{1}}\hat{p}^{\overline{2}}\cdots\hat{p}^{\overline{n}}
\right)\vert 0\rangle.
\label{pricminmap}
\eeq
\noindent
In this way we have embedded an $n$-qubit state to $\mathcal{F}^{(n)}$.

Now we consider the fermionic SLOCC transformations of
Eqs.(\ref{valodislocc}) and (\ref{fertraf}). These are
transformations, characterized by a $2n\times 2n$ matrix
$\mathcal{S}=e^A$, that leave the $n$-particle subspace of the
Fock space invariant. Our aim is to restrict $\mathcal{S}$ in such
a way that the resulting matrix also leaves the $n$-qubit
subspace, spanned by the basis vectors of Eq.(\ref{nqubsub}),
invariant and at the same time this new matrix also gives rise to
the usual $SL(2,\mathbb{C})^{\times n}$ part of the SLOCC action
on $\vert Z_{\psi}\rangle$.

Looking at Eq.(\ref{fertrafn}) it is easy to see that such
transformations can be organized to a matrix of the form \beq
{S}=\begin{pmatrix}a&b\\c&d\end{pmatrix}\in SL(2n,\mathbb{C})
\label{nq1} \eeq \noindent where $a=diag(a_1,\dots a_n)$,
$b=diag(b_1,\dots b_n)$, $c=diag(c_1,\dots c_n)$,
$diag(d_1,\dots d_n)$, i.e. the $n\times n$ blocks of $S$ are
{\it diagonal matrices}. One can also place these complex numbers
into an $n$ element set of $2\times 2$ matrices \beq
S^{(l)}\equiv\begin{pmatrix}a_l&b_l\\c_l&d_l\end{pmatrix}\in
SL(2,\mathbb{C}),\quad l=1,2,\dots n. \eeq \noindent Taken
together with the extra factor $\mathbb{C}^{\times}$ known from
Eq.(\ref{genslocc}) the transformation $\vert
Z_{\psi}\rangle\mapsto \lambda\hat{S}\vert Z_{\psi}\rangle$ gives
rise to the one \beq \psi_{\mu_1\dots\mu_n}\mapsto
{{\mathcal{A}^{(1)}}_{\mu_1}}^{\nu_1} \cdots
{{\mathcal{A}^{(n)}}_{\mu_n}}^{\nu_n}\psi_{\nu_1\dots\nu_n},\qquad
\mathcal{A}^{(l)}\in GL(2,\mathbb{C}),\quad l=1,2\dots n \eeq
\noindent which is just the usual SLOCC action for $n$-qubits.

In the case of SLOCC classification one generally obtains different {\it families} of entangled states. The families can contain inequivalent orbits under the SLOCC group. It can also happen that under permutations of the qubits one particular orbit
in a family is mapped to another orbit in another family.
Then in the case of qubits it is rewarding to explore the orbits of the group
$\tilde{G}=S_n\ltimes G$ where $G=GL(2,\mathbb{C})^{\times n}$ and $S_n$ is the symmetric group.
One can then ask what is the relationship between the {\it largest subgroup} $G^{\prime}$ of the fermionic SLOCC group $GL(2n,\mathbb{C})$ which leaves invariant the $n$-qubit subspace spanned by the vectors of Eq.(\ref{nqubsub}), and $\tilde{G}$. Suprisingly according to Lemma III.8. of Ref.\cite{Oeding} the answer to this question is $G^{\prime}=\tilde{G}$.
For this to make sense one should embed $S_n$ into $GL(2n,\mathbb{C})$ such that for an element $\sigma\in S_n$ we have
\beq
(1,2,\dots, n,\overline{1},\overline{2},\dots,\overline{n})\mapsto
(\sigma(1),\sigma(2),\dots, \sigma(n),\sigma(\overline{1}),\sigma(\overline{2}),\dots,\sigma(\overline{n})),
\eeq
\noindent
meaning that the basis vectors of Eq.(\ref{ujracim}) should be transformed accordingly.
The group $\tilde{G}$ will be used in Sec. 6.7. when studying four qubit invariants.

For illustrative purposes it is useful to invoke the following
physical interpretation\cite{Chen}. Our Hilbert space of one-particle states
is $\mathcal{H}=\mathbb{C}^{2n}\simeq \mathbb{C}^n\otimes
\mathbb{C}^2\equiv\mathcal{H}_{site}\otimes\mathcal{H}_{spin}$. In
this picture the fermions can be localized to $n$ sites (boxes),
and each site (box) can be filled with a spin which is either up
or down. This way of representing the $2^n$ basis states of
Eq.(\ref{nqubsub}) will be called {\it single occupancy}
representation. The remaining ${2n\choose n}-2^n$ basis states
contain {\it double and mixed occupancy states} as well. In this case some
of the boxes can also be empty or filled with two spins, one is up
the other is down.
The single and double occupancy representations of qubits are illustrated
in Figures 1. and 2.

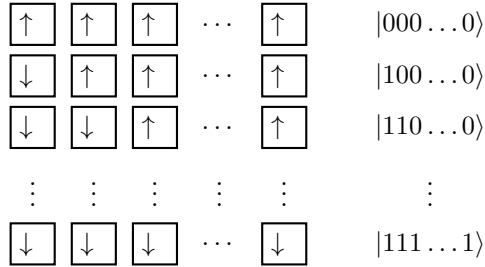
\begin{figure}[!h]
\begin{center}
\begin{tikzpicture}[scale=3]
    \tikzstyle{ann} = [draw=none,fill=none]
    \matrix[nodes={draw,  thick, fill=white!20},
        row sep=0.1cm,column sep=0.2cm] {
   &\ue;&\ue;&\ue; &\node[ann] {$\dots$};  &\ue    ; &\node[ann] {~~~~~$|000\dots0\rangle$};  \\
     &\de;&\ue;&\ue; &\node[ann] {$\dots$};  &\ue    ;  &\node[ann] {~~~~~$|100\dots0\rangle$};   \\
       &\de;&\de;&\ue; &\node[ann] {$\dots$};  &\ue    ;  &\node[ann] {~~~~~$|110\dots0\rangle$};  \\
  &\node[ann] {$\vdots$}; &\node[ann] {$\vdots$}; &\node[ann] {$\vdots$}; &\node[ann] {$\vdots$};&\node[ann] {$\vdots$};&\node[ann] {~~~~~$\vdots$};\\
       &\de;&\de;&\de; &\node[ann] {$\dots$};  &\de    ;  &\node[ann] {~~~~~$|111\dots1\rangle$};   \\
    };

\end{tikzpicture}
\end{center}
\caption{Single occupancy embedding of the $n$-qubit Hilbert space ($2^n$ basis vectors) inside $\mathcal{F}_{+}$ (for $n=2k$ boxes and $N=2n$ single particle states) or $\mathcal{F}_{-}$ (for $n=2k+1$ boxes and $N=2n$ single particle states)}
\end{figure}

\begin{figure}[!h]
\begin{center}
\begin{tikzpicture}[scale=3]
    \tikzstyle{ann} = [draw=none,fill=none]
    \matrix[nodes={draw,  thick, fill=white!20},
        row sep=0.1cm,column sep=0.2cm] {
     \ee;&\ee; &\ee; &\node[ann] {$\dots$};  &\ee       ; &\node[ann] {~~~~~$|000\dots0\rangle$}; \\
      \du;&\ee; &\ee; &\node[ann] {$\dots$};  &\ee     ;   &\node[ann] {~~~~~$|100\dots0\rangle$};  \\
       \du;&\du; &\ee; &\node[ann] {$\dots$};  &\ee      ;  &\node[ann] {~~~~~$|110\dots0\rangle$};   \\
       \node[ann] {$\vdots$}; &\node[ann] {$\vdots$}; &\node[ann] {$\vdots$};&\node[ann] {$\dots$}; &\node[ann] {$\vdots$};&\node[ann] {~~~~~$\vdots$}; \\
        \du;&\du; &\du; &\node[ann] {$\dots$};  &\du   ;  &\node[ann] {~~~~~$|111\dots1\rangle$};  \\
    };

\end{tikzpicture}
\end{center}
\caption{Double occupancy embedding of the $n$-qubit Hilbert space ($2^n$ basis vectors) inside $\mathcal{F}_+$ ($n$ boxes and $N=2n$ single particle states)}
\end{figure}
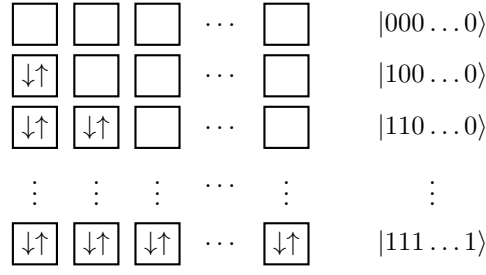

\bigskip
\section{Ways of embedding qubits}
\bigskip
\subsection{Embedding one qubit}
\bigskip
Apart from the "canonical" way discussed in the previous section,
there exist actually many more ways for obtaining embedded $n$-qubit systems.
Since these different types of embedding will be of
importance for us, here we start to clarify the technique of embedding.
We start with the elementary case of a single qubit.

In the case of a single qubit we have $n=1$ and
${V}=\mathbb{C}^2$. We have states from the even chirality sector
such as \beq \vert\psi_+\rangle =\left(\eta\hat{\bf
{1}}+\xi\hat{p}^1\hat{p}^{\overline{1}}\right)\vert
0\rangle\in\mathcal{F}_+ ,\qquad
\eta,\xi\in\mathbb{C}\label{plusch} \eeq \noindent and for the odd
chirality sector such as \beq
\vert\psi_-\rangle=\left(Z_1\hat{p}^1+Z_{\overline{1}}\hat{p}^{\overline{1}}\right)\vert
0\rangle\in\mathcal{F}_-,\qquad Z_1,Z_{\overline{1}}\in\mathbb{C}.
\label{negch} \eeq \noindent The nontrivial part of the
generalized SLOCC group comprises the group ${\rm
Spin}(4,\mathbb{C})$. According to Eq.(\ref{sexplicit}) an
element of this group can be written in
the form $\hat{S}=e^{\hat{s}}$ where \beq
\hat{s}={A_1}^1\hat{p}^1\hat{n}_1+
{A_{\overline{1}}}^{\overline{1}}\hat{p}^{\overline{1}}\hat{n}_{\overline{1}}+
{A_{\overline{1}}}^1\hat{p}^{\overline{1}}\hat{n}_1+
{A_1}^{\overline{1}}\hat{p}^{1}\hat{n}_{\overline{1}}+
B_{1\overline{1}}\hat{p}^1\hat{p}^{\overline{1}}+C^{1\overline{1}}\hat{n}_1\hat{n}_{\overline{1}}-\frac{1}{2}({A_1}^1+{A_{\overline{1}}}^{\overline{1}})\hat{\bf{1}}.
\label{1qs} \eeq \noindent Under
$\vert\psi\rangle\mapsto\hat{s}\vert\psi\rangle$ where
$\vert\psi\rangle=\vert\psi_+\rangle+\vert\psi_-\rangle$ we have
$(\eta,\xi)\mapsto (\eta^{\prime},\xi^{\prime})$ and
$(Z_1,Z_{\overline{1}})\mapsto
(Z^{\prime}_1,Z^{\prime}_{\overline{1}})$ where \beq
\begin{pmatrix}\eta^{\prime}\\\xi^{\prime}\end{pmatrix}=
\begin{pmatrix}-\frac{1}{2}\left({A_1}^1+{A_{\overline{1}}}^{\overline{1}}\right)&-C^{1\overline{1}}\\B_{1\overline{1}}&\frac{1}{2}\left({A_1}^1+{A_{\overline{1}}}^{\overline{1}}\right)\end{pmatrix}
\begin{pmatrix}\eta\\\xi\end{pmatrix}
\label{Arep}
\eeq
\noindent
\beq
\begin{pmatrix}Z^{\prime}_1\\Z^{\prime}_{\overline{1}}\end{pmatrix}=
\begin{pmatrix}\frac{1}{2}\left({A_1}^1-{A_{\overline{1}}}^{\overline{1}}\right)
&
{A_1}^{\overline{1}}    \\{A_{\overline{1}}}^1&
-\frac{1}{2}\left({A_1}^1-{A_{\overline{1}}}^{\overline{1}}\right)\end{pmatrix}
\begin{pmatrix}Z_1\\Z_{\overline{1}}\end{pmatrix}.
\label{Brep}
\eeq
\noindent

It is obvious that the state $\vert\psi_-\rangle$ is an embedded qubit. Its embedding is described by the usual process based on Eq.(\ref{pricminmap}).
Looking at (\ref{Brep}) we see that in this case the one parameter subgroups of $SL(2,\mathbb{C})$ are
\beq
\begin{pmatrix}\alpha&0\\0&\frac{1}{\alpha}\end{pmatrix},\quad
\begin{pmatrix}1&0\\
\beta&0\end{pmatrix},\quad
\begin{pmatrix}1&\gamma\\
0&1\end{pmatrix},\qquad
\log\alpha\equiv\frac{1}{2}({A_1}^1-{A_{\overline{1}}}^{\overline{1}}),\quad \beta\equiv {A_{\overline{1}}}^1,\quad \gamma\equiv{A_1}^{\overline{1}}.
\label{Btraf}
\eeq
\noindent

On the other hand due to its transformation properties under $SL(2,\mathbb{C})$
the state $\vert\psi_+\rangle$ is also a qubit. However, it is an unusual one. Its state space is a subspace of $\mathcal{F}$ where the particle number is {\it not conserved}.
Indeed, according to Eq.(\ref{sexplicit}) the SLOCC transformations also contain the transformations $e^{\hat{B}}$, the so called $B$-transforms that are creating two particles from the vacuum.
Similarly, we have $e^{\hat{C}}$, the $C$-transforms that are annihilating two particles from a  two particle state.
The corresponding one parameter subgroups are
\beq
\begin{pmatrix}a&0\\0&\frac{1}{a}\end{pmatrix},\quad
\begin{pmatrix}1&0\\
b&0\end{pmatrix},\quad
\begin{pmatrix}1&c\\
0&1\end{pmatrix},\qquad \log
a\equiv-\frac{1}{2}({A_1}^1+{A_{\overline{1}}}^{\overline{1}}),
\quad b\equiv B_{1\overline{1}},\quad c\equiv -C^{1\overline{1}}.
\label{Atraf}
\eeq
\noindent The four dimensional space $\mathcal{F}$ is a {\it
direct sum}: $\mathcal{F}=\mathcal{F}_+\oplus\mathcal{F}_-$. Let
us call the restriction of $\hat{s}$ of Eq.(\ref{1qs}) with
${A_1}^1=-{A_{\overline{1}}}^{\overline{1}}$ and
$B_{1\overline{1}}={C}^{1\overline{1}}=0$ the operator $\hat{s}_-$
\beq
\hat{s}_-=\log\alpha\left(\hat{p}^1\hat{n}_1-\hat{p}^{\overline{1}}
\hat{n}_{\overline{1}}\right)+\beta\hat{p}^{\overline{1}}n_1+\gamma\hat{p}^1\hat{n}_{\overline{1}}.
\label{esminusz} \eeq \noindent Similarly the restriction with
${A_1}^1={A_{\overline{1}}}^{\overline{1}}$ and
${A_1}^{\overline{1}}= {A_{\overline{1}}}^1=0$ will be called
$\hat{s}_+$ \beq \hat{s}_+=\log
a\left(\hat{n}_1\hat{p}^1-\hat{p}^{\overline{1}}
\hat{n}_{\overline{1}}\right)+b\hat{p}^1\hat{p}^{\overline{1}}-c\hat{n}_1\hat{n}_{\overline{1}}.
\label{esplusz} \eeq \noindent Then ${\hat
s}=\hat{s}_++\hat{s}_-$. This corresponds to the well-known fact
that $\mathfrak{s}\mathfrak{p}\mathfrak{i}\mathfrak{n}(4)\equiv \mathfrak{s}\mathfrak{l}(2)\oplus \mathfrak{s}\mathfrak{l}(2)$. Notice that due to
$\hat{s}_{\pm}\vert\psi_{\mp}\rangle =0$ we have
$\hat{s}\vert\psi\rangle=\hat{s}_+\vert\psi_+\rangle+\hat{s}_-\vert\psi_-\rangle$.

Let us relate the two different realizations of qubits on $\mathcal{F}$.
Recall the operators of Eq.(\ref{gammam}).
Clearly
\beq
\hat{\Gamma}_1\vert 0\rangle=\hat{p}^1\vert 0\rangle,\qquad
\hat{\Gamma}_1\hat{p}^1\hat{p}^{\overline{1}}\vert 0\rangle=\hat{p}^{\overline{1}}\vert 0\rangle
\label{basicmap}
\eeq
\noindent
Since $\hat{\Gamma}_1^2=\hat{\bf{1}}$ one can move between the basis vectors of the realizations of Eq.(\ref{plusch})-(\ref{negch}) back and forth.
Moreover, if $(\alpha,\beta,\gamma)\mapsto (a,b,c)$ then
\beq
\hat{\Gamma}_1\hat{s}_-\hat{\Gamma}_1=\hat{s}_+.
\label{kapcsqub}
\eeq
\noindent

\subsection{Embedding two qubits}
\bigskip
Though it was useful for setting the stage,
the previous case was physically unintresting.
This case was lacking the phenomenon of entanglement our main concern.
Now for the problem of embedding entangled qubits in different ways we consider our first nontrivial example, the case of two qubits.
We have $n=2$ and $N=4$ hence the generalized SLOCC group is $\mathbb{C}^{\times}\times {\rm Spin}(8,\mathbb{C})$.
In this case we have the range of indices $i,j=1,2,\overline{1},\overline{2}$ and the parametrizations
\beq
\vert\psi_+\rangle=\left(\eta\hat{\bf 1}+\frac{1}{2!}Z_{ij}\hat{p}^i\hat{p}^j
+\xi\hat{p}^1\hat{p}^{\overline{1}}\hat{p}^2\hat{p}^{\overline{2}}
\right)\vert 0\rangle,\qquad Z_{ij}=-Z_{ji}
\label{2qbitArep}
\eeq
\noindent
\beq
\vert\psi_-\rangle=\left(X_i\hat{p}^i+\frac{1}{3!}\epsilon_{ijkl}Y^i
\hat{p}^j\hat{p}^k\hat{p}^l\right)\vert 0\rangle.
\label{2qbitBrep}
\eeq
\noindent
Our aim is to identify two-qubit systems inside the $8$ dimensional Fock spaces
$\mathcal{F}_{\pm}$.
On each space $\mathcal{F}_{\pm}$ four copies of $SL(2,\mathbb{C})$ act.
Their $3\times 4=12$ complex parameters can be accomodated in a generator of the form Eq.(\ref{sexplicit}) with parameters placed inside the matrices $A,B,C$
as
\beq
{A_i}^j=
\begin{pmatrix}
\log\alpha_1-\log a_1&0&\gamma_1&0\\
0&\log\alpha_2-\log a_2&0&\gamma_2\\
\beta_1&0&-\log\alpha_1-\log a_1&0\\
0&\beta_2&0&-\log\alpha_2-\log a_2\end{pmatrix}
\label{Ama}
\eeq
\noindent
\beq
B_{ij}=\begin{pmatrix}0&0&b_1&0\\
0&0&0&b_2\\
-b_1&0&0&0\\
0&-b_2&0&0\end{pmatrix},\qquad
C^{ij}=\begin{pmatrix}0&0&-c_1&0\\
0&0&0&-c_2\\
c_1&0&0&0\\
0&c_2&0&0\end{pmatrix}.\label{BCma} \eeq \noindent Clearly inside
$\mathcal{F}_+$ we have a subsystem arising from the mapping of
Eq.(\ref{pricminmap}). This is the subsystem of single occupancy
states. States of this subsystem are of the form \beq \vert
\psi_+^{\rm{single}}\rangle\equiv \left(Z_{12}\hat{p}^1\hat{p}^{2}
+Z_{1\overline{2}}\hat{p}^1\hat{p}^{\overline{2}}+Z_{\overline{1}2}\hat{p}^{\overline{1}}\hat{p}^2+
Z_{\overline{1}\overline{2}}\hat{p}^{\overline{1}}\hat{p}^{\overline{2}}\right)\vert
0\rangle\in\mathcal{F}_+. \label{reszletes2q} \eeq \noindent On
this state two copies of $SL(2,\mathbb{C})$s act nontrivially.
Call
them $SL(2,\mathbb{C})_A$ and $SL(2,\mathbb{C})_C$. The
generators of these groups are of the same form as the right hand side
of Eq.(\ref{esminusz}),
where the range of indices is either $1,\overline{1}$ or
$2,\overline{2}$. The parameters are $(\alpha_1,\beta_1,\gamma_1)$
and $(\alpha_2,\beta_2,\gamma_2)$ respectively. We will call the
four dimensional subspace that the state of Eq.(\ref{reszletes2q}) belongs to as $V_{AC}$.
It is easy to check that the remaining two copies of
$SL(2,\mathbb{C})$s , to be called $SL(2,\mathbb{C})_B$ and
$SL(2,\mathbb{C})_D$ having the same form as the right hand side of Eq.(\ref{esplusz}) and
characterized by the parameters $(a_1,b_1,c_1)$ and
$(a_2,b_2,c_2)$, act on $V_{AC}$ trivially. This means that the
corresponding generators with the (\ref{esplusz}) form annihilate
$\vert \psi_+^{\rm{single}}\rangle$.
The single occupancy embedding of two qubits is illustrated in Figure 3.

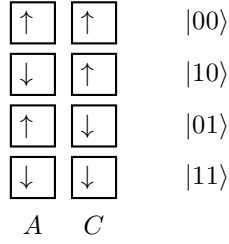
\begin{figure}[!h]
\begin{center}
\begin{tikzpicture}[scale=3]
    \tikzstyle{ann} = [draw=none,fill=none]
    \matrix[nodes={draw,  thick, fill=white!20},
        row sep=0.1cm,column sep=0.2cm] {
   &\ue;&\ue;   &\node[ann] {~~~~~$|00\rangle$}; \\
     &\de;&\ue; &\node[ann] {~~~~~$|10\rangle$};  \\
       &\ue;&\de;  &\node[ann] {~~~~~$|01\rangle$};  \\
       &\de;&\de;  &\node[ann] {~~~~~$|11\rangle$};  \\
       &\node[ann] {$A$}; &\node[ann] {$C$};\\
    };

\end{tikzpicture}
\end{center}
\caption{Single occupancy embedding of the $2$-qubit Hilbert space inside $\mathcal{F}_{+}$}
\end{figure}

Due to the product nature of the action of
$SL(2,\mathbb{C})_A\times SL(2,\mathbb{C})_C$ one can regard the space $V_{AC}$
as a one having a tensor product structure corresponding to two qubits, i.e. $V_{AC}=V_A\otimes V_C$.
Similar reasoning shows that the double occupancy subspace $V_{BD}$ with the representative
\beq
\vert \psi_+^{\rm{double}}\rangle=\left(\eta\hat{\bf{1}}+Z_{1\overline{1}}\hat{p}^1\hat{p}^{\overline{1}}
+Z_{2\overline{2}}\hat{p}^2\hat{p}^{\overline{2}}+\xi\hat{p}^1\hat{p}^{\overline{1}}\hat{p}^2\hat{p}^{\overline{2}}\right)\vert 0\rangle
\label{doubleocc}
\eeq
\noindent
is annihilated by
$SL(2,\mathbb{C})_A\times SL(2,\mathbb{C})_C$, but having a usual
$SL(2,\mathbb{C})_B\times SL(2,\mathbb{C})_D$ action.
Hence one can write $V_{BD}=V_B\otimes V_D$.
The double occupancy embedding of two qubits is illustrated in Figure 4.

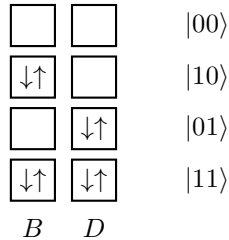
\begin{figure}[!h]
\begin{center}
\begin{tikzpicture}[scale=3]
    \tikzstyle{ann} = [draw=none,fill=none]
    \matrix[nodes={draw,  thick, fill=white!20},
        row sep=0.1cm,column sep=0.2cm] {
   &\ee;&\ee; &\node[ann] {~~~~~$|00\rangle$};   \\
     &\du;&\ee; &\node[ann] {~~~~~$|10\rangle$};  \\
       &\ee;&\du;  &\node[ann] {~~~~~$|01\rangle$};  \\
       &\du;&\du;  &\node[ann] {~~~~~$|11\rangle$};  \\
   &\node[ann] {$B$}; &\node[ann] {$D$};\\
   };

\end{tikzpicture}
\end{center}
\caption{Double occupancy embedding of the $2$-qubit Hilbert space inside $\mathcal{F}_{+}$}
\end{figure}

The result of these considerations is that one can write
\beq
\mathcal{F}_+=(V_A\otimes V_C)\oplus (V_B\otimes V_D)
\label{tensorsum1}
\eeq
\noindent
Of course this is just the well-known fact  that the spinor representation $8_s$ of ${\rm Spin}(8,\mathbb{C})$ under the subgroup $SL(2)_A\times SL(2)_B\times SL(2)_C\times SL(2)_D$ decomposes as
\beq
8_s=(2,1,2,1)\oplus (1,2,1,2).
\eeq
\noindent

It is important to realize at this point that the
(\ref{tensorsum1}) structure of $\mathcal{F}_+$ is induced by our
choice of the physically relevant subset of SLOCC transformations.
In particular the tensor product structures $V_{A}\otimes V_{C}$
and $V_{B}\otimes V_{D}$ are induced  by the input coming from
physics, namely our identification of a subset of transformations
playing a special role. Under this process we distinguished the
four generators $\hat{s}_A,\hat{s}_B,\hat{s}_C$ and $\hat{s}_D$ as
the ones representing a special set of physical protocols to be
performed on the physical states represented by elements of our
fermionic Fock space $\mathcal{F}_+$. In order to shed some light
on what do we mean by a "special set of physical protocols" let us
write out explicitly $\hat{s}_A,\hat{s}_B,\hat{s}_C$ and
$\hat{s}_D$ \beq
\hat{s}_A=\log\alpha_1\left(\hat{p}^1\hat{n}_1-\hat{p}^{\overline{1}}
\hat{n}_{\overline{1}}\right)+\beta_1\hat{p}^{\overline{1}}n_1+\gamma_1\hat{p}^1\hat{n}_{\overline{1}}
\label{sa} \eeq \noindent \beq \hat{s}_B=\log
a_1\left(\hat{n}_1\hat{p}^1-\hat{p}^{\overline{1}}
\hat{n}_{\overline{1}}\right)+b_1\hat{p}^1\hat{p}^{\overline{1}}-c_1
\hat{n}_1\hat{n}_{\overline{1}} \label{sb} \eeq \noindent \beq
\hat{s}_C=\log\alpha_2\left(\hat{p}^2\hat{n}_2-\hat{p}^{\overline{2}}
\hat{n}_{\overline{2}}\right)+\beta_2\hat{p}^{\overline{2}}n_2+\gamma_2\hat{p}^2\hat{n}_{\overline{2}}
\label{sc} \eeq \noindent \beq \hat{s}_D=\log
a_2\left(\hat{n}_2\hat{p}^2-\hat{p}^{\overline{2}}
\hat{n}_{\overline{2}}\right)+b_2\hat{p}^2\hat{p}^{\overline{2}}-c_2
\hat{n}_2\hat{n}_{\overline{2}}. \label{sd} \eeq \noindent From
these expressions it is clear that transformations $\hat{s}_{A,B}$
act on the modes $\{1,\overline{1}\}$, and $\hat{s}_{C,D}$ act on
the ones $\{2,\overline{2}\}$ of the  Hilbert space of
single-particle states. In the box picture these
operations act on the states of the first and second box
respectively. Moreover, the difference between $\hat{s}_{A,C}$ and
$\hat{s}_{B,D}$ is the one of single or double occupancy of the
corresponding box. When we think of the boxes as sites of a
lattice with two state systems (e.g. $1/2$ spins) attached to
them, the physical protocols are just the ones of addressing only
{\it one} of the sites and at the {\it same time also} deciding on
the (single or double occupancy) type of manipulations to be
performed on their spins. Clearly these types of manipulations
will provide different types of access to the resources available
in this simple lattice system characterized by the spinor
$\vert\psi_+\rangle$.

Notice also that apart from the tensor product structures Eq.(\ref{tensorsum1}) is also featuring a {\it direct sum}. The two parts of this direct sum correspond to
the physical sectors of single or double occupancy. These sectors are reminiscent of some superselection sectors used in quantum theory.
Namely, if for some physical reason we have no access to physical manipulations represented by generalized SLOCC transformations intertwining between these sectors, then we say that a superselection rule forbids us to go from single to double occupancy or vice cersa.

A comment here is in order. It is important to realize that had we
immediately started with four qubits and the corresponding spaces
$V_A$, $V_B$, $V_C$ and $V_D$, {\it physically} we would have had
no {\it a priori} reason
 for using a
{\it mathematical} construct such as $V_{AC}\oplus V_{BD}$ for
quantum information processing. The reason is that in this case this construct is
not representing any physically sound entangled system.

Now thanks to our constructions based on fermionic Fock space the
status of {\it certain}\footnote{Of course we are not expecting
that {\it any} direct sum structure can be embedded into {\it
some} fermionic Fock space. For example in the conclusions we will see that though the $56$
dimensional fundamental representation  space of the exceptional group $E_7(\mathbb{C})$ is arising as a special direct sum of seven three-qubit sectors however, this structure cannot be embedded into a single fermionic Fock
space.
}
direct sums combined with tensor products has changed. Indeed, for
fermionic systems we have a sound generalization of the notion of
SLOCC transformations hence in this special case it is easy to
make {\it physical} sense of their embedded subsystems.

Closing this section let us also comment on a dual construction based on the odd chirality sector
$\mathcal{F}_-$.
Let us write Eq.(\ref{2qbitBrep}) in the form
\beq
\vert\psi_-\rangle=\vert\psi_-^{sd}\rangle+\vert\psi_-^{ds}\rangle
\label{felir}
\eeq
\noindent
where
\beq
\vert\psi_-^{sd}\rangle=\left(X_1\hat{p}^1+X_{\overline{1}}\hat{p}^{\overline{1}}+Y^{\overline{1}}\hat{p}^1\hat{p}^{2\overline{2}}-Y^1\hat{p}^{\overline{1}}
\hat{p}^{2\overline{2}}\right)\vert 0\rangle
\label{sdo}
\eeq
\noindent
\beq
\vert\psi_-^{ds}\rangle=\left(X_2\hat{p}^2+X_{\overline{2}}\hat{p}^{\overline{2}}+Y^{\overline{2}}\hat{p}^{1\overline{1}}
\hat{p}^2
-Y^2
\hat{p}^{1\overline{1}}
\hat{p}^{\overline{2}}
\right)\vert 0\rangle.
\label{dos}
\eeq
\noindent
Here we have employed the shorthand notation $\hat{p}^{1\overline{1}}\equiv\hat{p}^1\hat{p}^{\overline{1}}$ etc., moreover we have used the combinations of letters $sd$ and $ds$ to indicate the hybrid nature of these states, i.e. they are combinations like "single-double" or "double-single".
It means that
$\vert\psi_-^{sd}\rangle$ and $\vert\psi_-^{ds}\rangle$
represent
two qubit systems with one of the qubits is taken in single and the other in double occupancy representation.
The mixed occupancy embedding of two qubits corresponding to $\vert\psi_-^{sd}\rangle$
is illustrated in Figure 5.

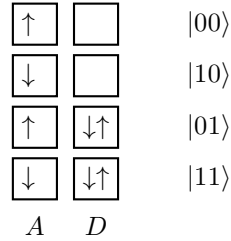
\begin{figure}[!h]
\begin{center}
\begin{tikzpicture}[scale=3]
    \tikzstyle{ann} = [draw=none,fill=none]
    \matrix[nodes={draw,  thick, fill=white!20},
        row sep=0.1cm,column sep=0.2cm] {
   &\ue;&\ee; &\node[ann] {~~~~~$|00\rangle$};   \\
     &\de;&\ee; &\node[ann] {~~~~~$|10\rangle$};  \\
       &\ue;&\du;  &\node[ann] {~~~~~$|01\rangle$};  \\
       &\de;&\du;  &\node[ann] {~~~~~$|11\rangle$};  \\
  &\node[ann] {$A$}; &\node[ann] {$D$};\\
  };

\end{tikzpicture}
\end{center}
\caption{Mixed occupancy embedding of the $2$-qubit Hilbert space inside $\mathcal{F}_{-}$}
\end{figure}

Now we have a decomposition
\beq
\mathcal{F}_-=\left(V_A\otimes V_D\right)\oplus\left(V_B\otimes V_C\right)
\label{masikdecomp}
\eeq
\noindent
which corresponds to the decomposition of the conjugate spinor representation $8_c$ of ${\rm Spin}(8,\mathbb{C})$ as
\beq
8_c=(2,1,1,2)\oplus (1,2,2,1).
\label{conjspin}
\eeq
\noindent

Let us write \beq
\mathcal{F}=\mathcal{F}_+\oplus\mathcal{F}_-=\left(\mathcal{F}^{00}\oplus\mathcal{F}^{11}\right)\oplus
\left(\mathcal{F}^{01}\oplus\mathcal{F}^{10}\right)=\left(V_{AC}\oplus
V_{BD}\right)\oplus \left(V_{AD}\oplus V_{BC}\right).
\label{qubitozas}\eeq \noindent This way of decomposing
$\mathcal{F}$ displays that qubits $A$ and $C$ are in the single
occupancy ($0$) and qubits $B$ and $D$ are in the double occupancy
($1$) representation.
 Notice that between the basis states of $V_{AC}$ and
$V_{BD}$ (i.e. $\mathcal{F}^{00}$ and $\mathcal{F}^{11}$ ) and the
action of the corresponding SLOCC groups $SL(2,\mathbb{C})_A\times
SL(2,\mathbb{C})_C$ and $SL(2,\mathbb{C})_B\times
SL(2,\mathbb{C})_D$ the operator $\hat{\Gamma}_1\hat{\Gamma}_2$
intertwine. Using similar intertwining properties one can write
\beq \mathcal{F}^{00}=\hat{\bf{1}}\mathcal{F}^{00},\qquad
\mathcal{F}^{10}=\hat{\Gamma}_1\mathcal{F}^{00},\qquad
\mathcal{F}^{01}=\hat{\Gamma}_2\mathcal{F}^{00},\qquad
\mathcal{F}^{11}=\hat{\Gamma}_1\hat{\Gamma}_2\mathcal{F}^{00}.
\label{intertwiners}\eeq \noindent Hence the state spaces of the
embedded two-qubit systems are generated from the one of the
canonical two-qubit system of Eq.(\ref{reszletes2q}) via the
action of suitable intertwining operators.

\subsection{Embedding three qubits}

Many aspects of the three qubit case have already been addressed within
a fermionic Fock space context\cite{SarLev1}.
However, related to the system of Eq.(\ref{pelda}) mentioned in the introduction it is important to revisit this case from the viewpoint of embedded systems.
Here we have $n=3$ and $N=6$ with the generalized SLOCC group $\mathbb{C}^{\times}\times{\rm Spin}(12,\mathbb{C})$.
Let us parametrize in this case the Weyl spinors of positive and negative chirality as
\beq
\vert\psi_-\rangle=\left(U_i\hat{p}^i+\frac{1}{3!}Z_{ijk}\hat{p}^{ijk}+\frac{1}{5!}W^i\varepsilon_{ijklmn}\hat{p}^{ijklm}\right)\vert 0\rangle
\label{3fermplusz}
\eeq
\beq
\vert\psi_+\rangle=
\left(\eta\hat{\bf {1}}+\frac{1}{2!}Y_{ij}\hat{p}^{ij}+\frac{1}{2!4!}X^{ij}\varepsilon_{ijklmn}\hat{p}^{klmn}+\xi\hat{p}^{123456}\right)\vert 0\rangle
\label{3fermminusz}
\eeq
\noindent
where $\hat{p}^{ijk}=\hat{p}^i\hat{p}^j\hat{p}^k$ etc.

Now we have $(1,2,3,4,5,6)\equiv(1,2,3,\overline{1},\overline{2},\overline{3})$. As usual
in the box picture we have three boxes or sites with two possible spin projections: up or down.
Now on each space $\mathcal{F}_{\pm}$ six copies of $SL(2,\mathbb{C})$ act.
Their $3\times 6=18$ parameters are placed inside the $6\times 6$ matrices $A,B,C$ similar to the pattern we already know form Eqs.(\ref{Ama})-(\ref{BCma}).

In the case of fermions with six single particle states the canonical three-qubit system connected to single occupancy is living inside $\mathcal{F}_-$.
It is related to the general pattern of embedding known from Eq.(\ref{pricminmap}).
Using the notation familiar from the end of the previous subsection we denote this subspace as $\mathcal{F}^{000}$. Hence we have
\beq
\vert\psi^{sss}_-\rangle\equiv\left(Z_{123}\hat{p}^{123}+Z_{12\overline{3}}\hat{p}^{12\overline{3}}+\cdots +Z_{\overline{123}}\hat{p}^{\overline{123}}\right)\vert 0\rangle\in\mathcal{F}^{000}.
\label{canon3qbi}
\eeq
\noindent
The notation "sss" or $000$ refers to the three $SL(2,\mathbb{C})$ generators that act nontrivially on this state. They are all in the single occupancy representation. This means that we have to use three copies of generators of the form of
Eqs.(\ref{sa}) ,(\ref{sc}) and a third one with labels featuring $3$ and $\overline{3}$. The remining three copies of $SL(2,\mathbb{C})$s with generators having the form of Eq.(\ref{sb}), (\ref{sd}) and again a third one act trivially on $\mathcal{F}^{000}$.
The result of these considerations is that now we have the decomposition of $\mathcal{F}$ to the
$32$ and $32^{\prime}$ representations corresponding to $\mathcal{F}_{\pm}$ as follows
\beq
\mathcal{F}=\mathcal{F}_+\oplus\mathcal{F}_-=\left(\mathcal{F}^{001}\oplus\mathcal{F}^{010}\oplus\mathcal{F}^{100}\oplus\mathcal{F}^{111}\right)\oplus
\left(\mathcal{F}^{000}\oplus\mathcal{F}^{011}\oplus\mathcal{F}^{101}\oplus\mathcal{F}^{110}\right).
\label{3qdekompozicio}
\eeq
\noindent
Clearly one can identify eight copies of three-qubit systems living inside $\mathcal{F}$. Unlike in the two-qubit case, now the single and double occupancy subspaces, namely $\mathcal{F}^{000}$ and $\mathcal{F}^{111}$ are living inside subspaces of different chirality.
The double occupancy state is of the form
\beq
\vert\psi^{ddd}_+\rangle=\left(\eta\hat{\bf {1}}+Y_{1\overline{1}}\hat{p}^{1\overline{1}}+\cdots-X^{1\overline{1}}\hat{p}^{2\overline{2}3\overline{3}}-\cdots
-\xi\hat{p}^{1\overline{1}2\overline{2}3\overline{3}}\right)\vert 0\rangle\in\mathcal{F}^{111}.
\label{dddstate}
\eeq
\noindent
The subspaces like $\mathcal{F}^{001}$ are in a mixed representation, meaning that two of the qubits are in the single and one of the qubits is in the double occupancy representation of the $SL(2,\mathbb{C})^{\times 6}$ subgroup.

As in the two-qubit case one can see that
\beq \mathcal{F}^{000}=\hat{\bf{1}}\mathcal{F}^{000},\quad
\mathcal{F}^{100}=\hat{\Gamma}_1\mathcal{F}^{000},\quad
\dots\qquad
\mathcal{F}^{110}=\hat{\Gamma}_1\hat{\Gamma}_2\mathcal{F}^{000}
,\quad
\mathcal{F}^{111}=\hat{\Gamma}_1\hat{\Gamma}_2\hat{\Gamma}_3\mathcal{F}^{000}.
\label{intertwiners3q}\eeq \noindent Hence the state spaces of the
embedded three-qubit systems are again generated from the one of the
canonical three-qubit subspaces via the
action of suitable intertwining operators.

Let us  associate to the first box (site) qubits A, and B. Qubit A is in single
and qubit B in double occupancy. Similarly to the second box we associate C and D, for the third box E and F.
Then the decomposition of Eq.(\ref{3qdekompozicio}) takes the form
\beq
\mathcal{F}=\left(V_{ACF}\oplus V_{ADE}\oplus V_{BCE}\oplus V_{BDF}\right)
\oplus\left(V_{ACE}\oplus V_{ADF}\oplus V_{BCF}\oplus V_{BDE}\right).
\label{dekika}
\eeq
\noindent
Notice the structure of either $\mathcal{F}_+$ or $\mathcal{F}_-$.
One can arrange the four summands to the vertices of the tetrahedron.
Then the six edges will correspond to the six common qubits.
Alternatively one can consider an incidence geometry consisting of four points
labelled by triples like $ACF,ADE,BCE,BDF$ and the lines by $A,B,C,D,E,F$.
Then e.g. points  $ACF$ and $ADE$ are connected by line $A$. etc.
This incidence structure coincides with the one of the complement of a line of the Fano plane.
We also note that the decomposition of Eq.(\ref{qubitozas})
is precisely the one of Eq.(\ref{pelda}) familiar from the introduction
mentioned in connection with the BHQC.
However, unlike in previous attempts now to such constructs a : quantum  information theoretic meaning
was given.

It is instructive to calculate $\vert\tilde{\psi}_+\rangle=\hat{\Gamma}_1\hat{\Gamma}_2\hat{\Gamma}_3\vert\psi_-\rangle$.
This gives the $32$ complex amplitudes of the positive chirality part parametrized by the $32$ complex amplitudes of the negative chirality one.
The result is
\beq
\tilde{X}^{ij}=\begin{pmatrix}0&U_3&-U_2&Z_{1\overline{2}\overline{3}}&
Z_{1\overline{3}\overline{1}}&Z_{1\overline{1}\overline{2}}
\\-U_3&0&U_1&Z_{2\overline{2}\overline{3}}&Z_{2\overline{3}\overline{1}}&
Z_{2\overline{1}\overline{2}}\\
U_2&-U_1&0&Z_{3\overline{2}\overline{3}}&Z_{3\overline{3}\overline{1}}&
Z_{3\overline{1}\overline{2}}\\
-Z_{1\overline{2}\overline{3}}&-Z_{2\overline{2}\overline{3}}&
-Z_{3\overline{2}\overline{3}}&0&W^{\overline{3}}&-W^{\overline{2}}\\
-Z_{1\overline{3}\overline{1}}&-Z_{2\overline{3}\overline{1}}&
-Z_{3\overline{3}\overline{1}}&-W^{\overline{3}}&0&W^{\overline{1}}\\
-Z_{1\overline{1}\overline{2}}&-Z_{2\overline{1}\overline{2}}&-Z_{3\overline{1}\overline{2}}&W^{\overline{2}}&-W^{\overline{1}}&0
\label{ixtilde}
\end{pmatrix},\qquad\tilde{\xi}=Z_{\overline{123}}
\eeq
\noindent
\beq
\tilde{Y}_{ij}=\begin{pmatrix}0&-W_3&W_2&-Z_{\overline{1}23}&
-Z_{\overline{2}23}&-Z_{\overline{3}23}
\\W_3&0&-W_1&-Z_{\overline{1}31}&
-Z_{\overline{2}31}&-Z_{\overline{3}31}\\
-W_2&W_1&0&-Z_{\overline{1}12}&-Z_{\overline{2}12}&
-Z_{\overline{3}12}\\
Z_{\overline{1}23}&Z_{\overline{1}31}&
Z_{\overline{1}12}&0&-U_{\overline{3}}&U_{\overline{2}}\\
Z_{\overline{2}23}&Z_{\overline{2}31}&
Z_{\overline{2}12}&U_{\overline{3}}&0&-U_{\overline{1}}\\
Z_{\overline{3}23}&Z_{\overline{3}31}&Z_{\overline{3}12}&-U_{\overline{2}}&U_{\overline{1}}&0
\label{iytilde}
\end{pmatrix},
\qquad\tilde{\eta}=-Z_{123}.
\eeq
\noindent
This dictionary provides an explicit form for $\vert\tilde{\psi}_+^{ddd}\rangle
=\hat{\Gamma}_1\hat{\Gamma}_2\hat{\Gamma}_3\vert\psi^{sss}_-\rangle$
hence for the intertwining map between $\mathcal{F}^{000}$ and $\mathcal{F}^{111}$, i.e. the map between the single and double occupancy representation of three-qubits.
Note that the intertwiner above has a special significance in string theory.
It is related to the so called mirror map which for toroidal compactifications is relating via T-duality the IIA and IIB duality frames of the relevant string theories.
Restricting attention to the subset of the $8$ amplitudes
$(Z_{123},Z_{12\overline{3}},\dots,Z_{\overline{123}})$ describing three-qubit states
we obtain a new labelling for three-qubits.
Originally this unusual representation of three-qubits  equivalent to our double occupancy
representation was the first to appear within the context of the BHQC\cite{BHQC}.

\subsection{Embedding four qubits.}

We have $n=4$ and $N=8$ and the generalized SLOCC
group is $\mathbb{C}^{\times}\times{\rm Spin}(16,\mathbb{C})$. A
Weyl spinor of positive chirality is now of the form \beq
\vert\psi_+\rangle=\left(\eta\hat{\bf
{1}}+\frac{1}{2!}X_{ij}\hat{p}^{ij}+\frac{1}{4!}Z_{ijkl}\hat{p}^{ijkl}+
\frac{1}{6!}\varepsilon_{ijklmnrs}Y^{ij}\hat{p}^{klmnrs}+
\xi\hat{p}^{12345678} \right)\vert 0\rangle \label{4qfermi} \eeq
\noindent As usual the most natural way of embedding four qubits
into $\mathcal{F}_+$ is via single occupancy \beq
\vert\psi^{ssss}_+\rangle=\left(Z_{1234}\hat{p}^{1234}+Z_{123\overline{4}}
\hat{p}^{123\overline{4}}+\cdots+Z_{\overline{123}4}p^{\overline{123}4}+Z_{\overline{1234}}
\hat{p}^{\overline{1234}}\right)\vert
0\rangle\in\mathcal{F}^{0000}.\label{4qssss} \eeq \noindent There
are $8$ different embedded $4$ qubit subspaces in $\mathcal{F}_+$.
These are \beq
\mathcal{F}^{\mu_1\mu_2\mu_3\mu_4}\equiv\hat{\Gamma}_1^{\mu_1}
\hat{\Gamma}_2^{\mu_2}\hat{\Gamma}_3^{\mu_3}
\hat{\Gamma}_4^{\mu_4}\mathcal{F}^{0000},\qquad
\mu_1+\mu_2+\mu_3+\mu_4\equiv 0,\qquad
\mu_1,\mu_2,\mu_3,\mu_4\in\mathbb{Z}_2. \label{8drb4q} \eeq
\noindent Similarly, we have $8$ further embeddings into
$\mathcal{F}_-$ with $\mu_1+\mu_2+\mu_3+\mu_4=1$. Hence one can
write \beq
\mathcal{F}=\bigoplus_{(\mu_1\mu_2\mu_3\mu_4)\in(\mathbb{Z}_2)^4}
\mathcal{F}^{\mu_1\mu_2\mu_3\mu_4}. \label{dekompoz4qb} \eeq
\noindent

\section{Invariants and covariants.}

\subsection{ The  invariant bilinear form}
\bigskip

We start by recapitulating some of the results of Ref.\cite{SarLev1}.
Let us consider a collection of $k$ elements $\{x_1,x_2,\dots
,x_k\}$ of the vector space $\mathcal{V}$. These give rise to a
set $\{\hat{x}_1,\hat{x}_2,\dots, \hat{x}_k\}$ of operators. We
multiply these operators and define a map, called the {\it
transposed} map as follows \beq
(\hat{x}_1\hat{x}_2\cdots\hat{x}_k)^T=\hat{x}_k\cdots
\hat{x}_2\hat{x}_1. \label{transposed} \eeq \noindent Consider now
the operator $\hat{n}_1\hat{n}_2\cdots\hat{n}_N$. This operator
annihilates all the terms from the expansion of Eq.(\ref{steexp}) except the
term from the one dimensional subspace of $\mathcal{F}$
corresponding to the $\wedge^NV^{\ast}$ part of the (\ref{grass}) Grassmann
algebra. It is just the subspace spanned by the basis vector \beq
\vert{\rm top}\rangle\equiv\hat{p}^1\hat{p}^2\cdots\hat{p}^N\vert
0\rangle\label{TOP} \eeq \noindent For this vector we have \beq
\hat{n}_1\hat{n}_2\cdots\hat{n}_N\vert{\rm top}\rangle=
(-1)^{\frac{N(N-1)}{2}}\hat{n}_N\cdots\hat{n}_2\hat{n}_1\hat{p}^1\hat{p}^2\cdots\hat{p}^N\vert
0\rangle=(-1)^{\frac{N(N-1)}{2}}\vert 0\rangle. \label{tophatas}
\eeq \noindent Let us now consider two elements of the fermionic
Fock space \beq \vert\psi\rangle =\hat{\Psi}\vert
0\rangle\in\mathcal{F},\qquad \vert\phi\rangle =\hat{\Phi}\vert
0\rangle \in\mathcal{F}. \label{elemek} \eeq \noindent Our aim is
to define a nondegenerate bilinear form \beq
(\cdot,\cdot):\mathcal{F}\times\mathcal{F}\to\mathbb{C}\label{bili1}
\eeq \noindent invariant under the nontrivial subgroup of the
generalized SLOCC group i.e. ${\rm Spin}(2N,\mathbb{C})$. We
define \beq (\psi,\phi)\vert 0\rangle\equiv
(-1)^{\frac{N(N-1)}{2}}(\hat{n}_1\hat{n}_2\cdots\hat{n}_N)\hat{\Psi}^T\hat{\Phi}\vert
0\rangle. \label{Mukai} \eeq \noindent By virtue of
Eq.(\ref{tophatas}) the meaning of the bilinear form is clear: it
picks out the complex coefficient of the "top" part of the state
$\hat{\Psi}^T\hat{\Phi}\vert 0\rangle$. From the definition it is
obvious that for any operator $\hat{O}$ we have \beq
(\psi,\hat{O}\phi)=(\hat{O}^T\psi,\phi).\label{transmeaning} \eeq
\noindent One can also check that \beq
(\psi,\phi)=(-1)^{\frac{N(N-1)}{2}}(\phi,\psi). \label{symmprop}
\eeq \noindent Hence this bilinear form is symmetric for
$N=0,1({\rm mod}4)$ and antisymmetric for $N=2,3({\rm mod}4)$.

Now a look at Eq.(\ref{sexplicit}) shows that the generators of ${\rm
Spin}(2N,\mathbb{C})$ satisfy $\hat{s}^T=-\hat{s}$. Combining this
with Eq.(\ref{transmeaning}) shows that \beq
(\psi,\hat{s}\phi)+(\hat{s}\psi,\phi)=0 \label{proofinv} \eeq
\noindent which demonstrates the invariance of our bilinear form
under the generalized SLOCC transformations of the form
$\hat{S}=e^{\hat{s}}$.

\subsection{Invariants and covariants for $N=4,6$}

For $N=2n$ we have embedded $n$-qubit systems. Our aim is to
construct the basic covariants and invariants of fermionic systems
and relate these quantities to the corresponding ones of qubits.

 Let us construct covariants using our bilinear form. The
simplest choices are \beq K_I\equiv(\psi,\hat{e}_I\psi),\qquad
K^I=g^{IJ}K_J=(\psi,\hat{e}^I\psi).\label{basiccov} \eeq \noindent
Using Eqs.(\ref{transmeaning}), (\ref{spindefi}) and
$\hat{s}^T=-\hat{s}$ we see that
 under the generalized SLOCC group
 $K_I$ transforms as \beq (\psi,\hat{e}_I\psi)\mapsto
 (\hat{S}\psi,\hat{e}_I\hat{S}\psi)=(\psi,\hat{S}^{-1}\hat{e}_I\hat{S}\psi)=
 (\psi,\hat{e}_J\psi)(\mathcal{S}^{-1})^J_I \label{alsotrafo} \eeq
 \noindent As a result of this and Eq.(\ref{matrica}) one has \beq K_I\mapsto
 K_J{(\mathcal{S}^{-1})^J}_I,\qquad K^I\mapsto
      {\mathcal{S}^I}_JK^J.\label{covtr1}
           \eeq
            \noindent

        In order to build invariants one can try to experiment with
        further covariants. A natural choice is a covariant \beq
        {\mathcal{K}^I}_J\equiv
        (\psi,\hat{e}^I\hat{e}_{J}\psi).\label{momentmap} \eeq \noindent
        Since it transformas as
$\mathcal{K}\mapsto\mathcal{S}K\mathcal{S}^{-1}$ one can form the
invariants \beq I_{2n}\equiv {\rm Tr}(\mathcal{K}^n).
\label{mominvar} \eeq \noindent In order to explore the structure
of these invariants we write $\mathcal{K}$ as \beq
{\mathcal{K}^I}_J=\frac{1}{2}g^{IL}(\psi,[\hat{e}_L,\hat{e}_J]\psi)+
\frac{1}{2}g^{IL}(\psi,\{\hat{e}_L,\hat{e}_J\}\psi)=
\frac{1}{2}g^{IL}(\psi,[\hat{e}_L,\hat{e}_J]\psi)+\frac{1}{2}{\delta^I}_J(\psi,\psi)
\label{szamol1}
 \eeq \noindent
 where we have used $\{\hat{e}_I,\hat{e}_J\}=g_{IJ}\hat{1}$. 
 
 For
 $N=0,1({\rm mod} 4)$ the bilinear form is symmetric. In this case by virtue
 of the fact that the $[\hat{e}_I,\hat{e}_J]$ are just the
 generators of ${\rm Spin}(2N,\mathbb{C})$ and Eqs.(\ref{symmprop})
 the first term gives zero. Hence in this case \beq
 {\mathcal{K}^I}_J=\frac{1}{2}{\delta^I}_J(\psi,\psi).\label{cov4m}
 \eeq \noindent Using (\ref{mominvar}) we get \beq
 I_{2r}=2^{1-r}N(\psi,\psi)^r.\label{invar4m} \eeq \noindent Hence
 in this case apart from the quadratic invariant $(\psi,\psi)$ no
 new invariant of this kind is obtained.
We note that for the $r=1$, $N=4$ case a restriction of Eq.(\ref{2qbitArep})
to two fermions with four modes gives for $I_2$ four times a quadratic form which
corresponds to the usual Pl\"ucker relations. Its square is just the determinant of the $4\times 4$
antisymmetric matrix $Z_{ij}$.
The magnitude of this quadratic form up to constant factors is just the usual measure of entanglement introduced in\cite{Schliemann}, which for embedded two qubits boils down to the well-known concurrence.

 On the other hand for $N=2,3 ({\rm mod} 4)$ the bilinear
 form is antisymmetric hence the last term of (\ref{szamol1})
 vanishes giving the result \beq
 {\mathcal{K}^I}_J=\frac{1}{2}g^{IL}(\psi,[\hat{e}_L,\hat{e}_J]\psi).
 \label{cov4m2} \eeq \noindent Clearly since $g^{IJ}$ is symmetric
 and the commutator is antisymmetric in this case $I_{2}=0$. So the
 first new nontrivial nonzero invariant should be a quartic one,
 $I_4$.
 Indeed, a calculation in the special case $N=6$
 shows\cite{SarLev1,HL} that when restricted to the subspace of positive chirality
 this invariant is just the quartic invariant introduced by Igusa\cite{Igusa}
 for his classification of spinors up to $N=6$. This invariant is
 also related to the so called generalized Hitchin functional\cite{LH}.

 Notice also that in the case $N=4m+2$ using the matrices of
 Eq.(\ref{kompaktalak}) and (\ref{para}) one can form the new matrix $\Lambda\equiv sg$
 satisfying ${\Lambda}^t=-{\Lambda}^t$
  Then using Eq.(\ref{cov4m2}) one gets
\beq \frac{1}{2}{\rm
Tr}(s\mathcal{K}_{\psi})=(\psi,\hat{s}\psi),\qquad
\hat{s}=\frac{1}{2}\Lambda^{IJ}\hat{e}_I\hat{e}_J\in
\mathfrak{s}\mathfrak{p}\mathfrak{i}\mathfrak{n}(2N), \quad s\in \mathfrak{s}
\mathfrak{o}(2N) \label{alternative} \eeq
\noindent where one can check that this expression for $\hat{s}$
coincides with the usual one of Eq.(\ref{sexplicit}). Then we have a mapping $
\mathcal{F}_{\pm}\to {\rm so}(2N)$ of the form
$\vert\psi\rangle\mapsto\mathcal{K}_{\psi}$. Since in this case the (\ref{Mukai})
bilinear
form is antisymmetric, we can regard it as a symplectic form on
$\mathcal{F}_{\pm}$, hence we can think the spaces
$\mathcal{F}_{\pm}$ as phase spaces of a classical mechanical
system with the generalized SLOCC transformations defining a group action
on it. It can then be shown that in this case the association
$\vert\psi\rangle\mapsto\mathcal{K}_{\psi}$ described by
Eq.(\ref{alternative}) is the so called moment map \cite{Hitchin,
LH,SarLev1}.

\subsection{Invariants and covariants for $N=8$.}

Let us consider the case when $N\equiv 0({\rm mod}4)$. Here an
important special case is the $N=8$ one which contains four
fermions with eight single particle states.
This is the setting where one can embed naturally four-qubit systems
our main concern here.
The corresponding
state is living inside a Weyl spinor of the (\ref{4qfermi}) form.
As described in Eq.(\ref{dekompoz4qb}) this case
incorporates $8$ different classes of embedded four-qubit states.
The most natural embedding is the (\ref{4qssss}) one based on
single occupancy.

The basic covariant we should consider here is the one \beq
{\mathcal{K}^{IJ}}_{KL}\equiv
(\psi,\hat{e}^I\hat{e}^J\hat{e}_K\hat{e}_L\psi).
\label{Mkovarians} \eeq \noindent Writing
$\hat{e}^I\hat{e}^J=\frac{1}{2}[\hat{e}^I,\hat{e}^J]+\frac{1}{2}\{\hat{e}^I,\hat{e}^J\}$
and noting that in the $N\equiv 0({\rm mod}4)$ case
$(\psi,[\hat{e}_I,\hat{e}_J]\psi)=0$ and $(\psi,\phi)=(\phi,\psi)$
we get \beq {\mathcal{K}^{IJ}}_{KL}\equiv
g^{II^{\prime}}g^{JJ^{\prime}}(\psi,
\hat{e}_{I^{\prime}}\hat{e}_{J^{\prime}}\hat{e}_K\hat{e}_L\psi)=
{\mathcal{R}^{IJ}}_{KL}+\frac{1}{4}g^{IJ}g_{KL}(\psi,\psi)
\label{kov4ex1} \eeq \noindent where \beq {\mathcal{R}^{
IJ}}_{KL}=\frac{1}{4} g^{II^{\prime}}g^{JJ^{\prime}} (\psi,
[\hat{e}_{I^{\prime}},\hat{e}_{J^{\prime}}][\hat{e}_K,\hat{e}_L]\psi).
\label{kov4ex2} \eeq \noindent

For fermionic systems described by Weyl spinors of the form
(\ref{4qfermi}) the basic invariants under the generalized SLOCC
group ${\rm Spin}(16,\mathbb{C})$ are the ones \beq
\mathcal{J}_{2p}={\mathcal{K}^{I_1J_1}}_{I_2J_2}
{\mathcal{K}^{I_2J_2}}_{I_3J_3}\cdots
{\mathcal{K}^{I_pJ_p}}_{I_1J_1}. \label{gensloccN8} \eeq \noindent
However, due to Eq.(\ref{kov4ex1}) it is enough to consider the
invariants \beq \mathcal{I}_{2p}={\mathcal{R}^{I_1J_1}}_{I_2J_2}
{\mathcal{R}^{I_2J_2}}_{I_3J_3}\cdots
{\mathcal{R}^{I_pJ_p}}_{I_1J_1}. \label{eztitteleg} \eeq \noindent

\subsection{Four fermions with eight modes}

In the following we consider the positive chirality sector of the
$N=8$ case with $\vert\psi\rangle\equiv\vert\psi_+\rangle$ (see
Eq.(\ref{4qfermi})) with the constraint  \beq
\vert\psi\rangle=\frac{1}{4!}Z_{ijkl}\hat{p}^{ijkl}\vert 0\rangle.
\label{4fermi8mode} \eeq \noindent Now the quadratic SLOCC
invariant is
 \beq
(\psi,\psi)=\frac{1}{4!4!}\varepsilon^{ijklmnrs}Z_{ijkl}Z_{mnrs}.
\label{invkvad4} \eeq \noindent Recall now that
$\hat{e}_I=(\hat{e}_i,\hat{e}_{i+8})=(\hat{n}_i,\hat{p}^i)$ to
show that the only nonzero independent components of
${\mathcal{R}^{ IJ}}_{KL}$ are ${\mathcal{R}^{ ij+8}}_{k+8l}$ and
${\mathcal{R}^{ ij}}_{kl}$. For example we have \beq
{\mathcal{R}^{
ij}}_{kl}=(\psi,\hat{p}^{ij}\hat{n}_{kl}\psi)=(\psi,\hat{n}_{kl}\hat{p}^{ij}\psi)={\mathcal{R}^{
k+8l+8}}_{i+8j+8}.\label{katamatrix} \eeq \noindent Similarly we
have \beq {\mathcal{R}^{
ij+8}}_{k+8l}=\frac{1}{4}(\psi,[\hat{p}^i,\hat{n}_j][\hat{p}^k,
\hat{n}_l]\psi)= \frac{1}{4}(\psi,[\hat{p}^k,\hat{n}_l][\hat{p}^i,
\hat{n}_j]\psi)={\mathcal{R}^{kl+8}}_{i+8j}. \eeq \noindent For
the explicit form of  ${\mathcal{R}^{ij}}_{kl}$ one gets \beq
{\mathcal{R}^{
ij}}_{kl}=\frac{1}{2!4!}\varepsilon^{ijabcdef}Z_{lkab}Z_{cdef}.
\label{katanovasmatrix} \eeq \noindent On the other hand \beq
{\mathcal{R}^{
ij+8}}_{k+8l}=(\psi,\hat{p}^i\hat{n}_j\hat{p}^k\hat{n}_l\psi)+
\frac{1}{4}\delta^i_j\delta^k_l(\psi,\psi)-\frac{1}{2}\delta^k_l(\psi,\hat{p}^i\hat{n}_j\psi)-
\frac{1}{2}\delta^i_j(\psi,\hat{p}^k\hat{n}_l\psi). \eeq \noindent
Now \beq
(\psi,\hat{p}^i\hat{n}_j\psi)=\frac{1}{2}(\psi,[\hat{p}^i,\hat{n}_j]\psi)
+\frac{1}{2}(\psi,\{\hat{p}^i,\hat{n}_j\}\psi)=0+\frac{1}{2}\delta^i_j(\psi,\psi),
\eeq \noindent hence \beq {\mathcal{R}^{
ij+8}}_{k+8l}=(\psi,\hat{p}^i\hat{n}_j\hat{p}^k\hat{n}_l\psi)-
\frac{1}{4}\delta^i_j\delta^k_l(\psi,\psi) \eeq \noindent yielding
the result \beq {\mathcal{R}^{
ij+8}}_{k+8l}=\left(\frac{1}{2}\delta^k_j\delta^i_l-\frac{1}{4}\delta^i_j\delta^k_l\right)(\psi,\psi)-
{\mathcal{R}^{ik}}_{jl}. \label{katamatr} \eeq \noindent Then
according to Eq.(\ref{kov4ex1}) the net result is that all the
components of the covariant ${\mathcal{K}^{IJ}}_{KL}$ can entirely
be expressed in terms of the invariant $(\psi,\psi)$ and the
quantity ${\mathcal{R}^{ij}}_{kl}$ \beq
{\mathcal{K}^{ij}}_{kl}={\mathcal{R}^{ij}}_{kl}, \qquad
{\mathcal{K}^{ij+8}}_{k+8l}=\frac{1}{2}(\psi,\psi)\delta^i_l\delta^k_j-{\mathcal{R}^{
ik}}_{jl}. \label{Katanovavegso} \eeq \noindent
${\mathcal{R}^{ij}}_{kl}$ with the (\ref{katanovasmatrix})
explicit from is a covariant with respect to the ordinary SLOCC
subgroup $SL(8,\mathbb{C})$ of ${\rm Spin}(16,\mathbb{C})$. It is
just the covariant introduced by Katanova\cite{Katanova}. Hence
for the construction of invariants for embedded four fermionic
systems
 with eight single particle states it is enough to consider the invariants formed by the matrix
${\mathcal{R}^{ij}}_{kl}$ of Eq.(\ref{katanovasmatrix}). These
invariants are of the form
 \beq
I_{2p}={\mathcal{R}^{i_1j_1}}_{i_2j_2}
{\mathcal{R}^{i_2j_2}}_{i_3j_3}\cdots{\mathcal{R}^{i_pj_p}}_{i_1j_1}
. \label{fermiinvi} \eeq \noindent It is known\cite{Katanova,Chen}
that  \beq \{I_2,I_6,I_8,I_{10},I_{12}, I_{14},I_{18}\}
\label{algindepsetkata} \eeq \noindent gives an algebraically
independent set of generators.

\subsection{Embedded four-qubits, covariants and density matrices}

In this section via restricting the spinor of
Eq.(\ref{4fermi8mode}) we start deriving the usual set of four-qubit
invariants in a spinorial language. Our starting point is the
embedded four-qubit state in the single occupancy representation
\beq
\vert\psi\rangle=\left(\psi_{0000}\hat{p}^{1234}+\psi_{0001}\hat{p}^{123\overline{4}}+\dots
+\psi_{1110}\hat{p}^{\overline{1}\overline{2}\overline{3}4}+\psi_{1111}
\hat{p}^{\overline{1}\overline{2}\overline{3}\overline{4}}\right)\vert
0\rangle \in{\mathcal{F}}_+ \label{ezitta4qbit}\eeq \noindent
hence $\psi_{0000}=Z_{1234}$,
$\psi_{0001}=Z_{123\overline{4}}=Z_{1238}$ etc. This is to be
compared with the conventional way of writing this state as \beq
\vert\psi\rangle=\sum_{\mu_1\mu_2\mu_3\mu_4\in
0,1}\psi_{\mu_1\mu_2\mu_3\mu_4}\vert
\mu_1\mu_2\mu_3\mu_4\rangle\in {\mathbb{C}}^2\otimes
{\mathbb{C}}^2\otimes{\mathbb{C}}^2\otimes{\mathbb{C}}^2.
\label{usual4repr} \eeq \noindent Using the $16$ complex
amplitudes of our state one can define three basic $4\times 4$
matrices containing four four-component vectors $U,V,W$ and $Z$
\beq \label{Lmatrix4qbit} \mathcal{ L}\equiv
\begin {pmatrix}{\psi}_{0000}&{\psi}_{0001}&{\psi}_{0010}&{\psi}_{0011}\\
{\psi}_{0100}&{\psi}_{0101}&{\psi}_{0110}&{\psi}_{0111}\\
{\psi}_{1000}&{\psi}_{1001}&{\psi}_{1010}&{\psi}_{1011}\\
{\psi}_{1100}&{\psi}_{1101}&{\psi}_{1110}&{\psi}_{1111}\end{pmatrix}
\equiv
\begin{pmatrix}U^0&U^1&U^2&U^3\\
V^0&V^1&V^2&V^3\\
W^0&W^1&W^2&W^3\\
Z^0&Z^1&Z^2&Z^3\end{pmatrix} \eeq \noindent \beq
\label{MNmatrixok4qbit}
\mathcal{ M}=\begin{pmatrix}U^0&U^1&V^0&V^1\\
W^0&W^1&Z^0&Z^1\\
U^2&U^3&V^2&V^3\\
W^2&W^3&Z^2&Z^3\end{pmatrix},\qquad \mathcal{ N}=
\begin{pmatrix}U^0&U^2&V^0&V^2\\
U^1&U^3&V^1&V^3\\
W^0&W^2&Z^0&Z^2\\
W^1&W^3&Z^1&Z^3\end{pmatrix}. \eeq \noindent Notice that the
matrices $\mathcal{ M}$ and $\mathcal{ N}$ are obtained from the
matrix $\mathcal{ L}$ via the permutations.
The index structure of these matrices is \beq
\label{4qbitpermutaciok}\mathcal{ L}\leftrightarrow
{\psi}_{\mu_1\mu_2\mu_3\mu_4}\qquad \mathcal{ M}\leftrightarrow
{\psi}_{\mu_3\mu_1\mu_2\mu_4}\qquad\ \mathcal{ N}\leftrightarrow
{\psi}_{\mu_1\mu_4\mu_2\mu_3}.\eeq\noindent If we use the
(\ref{usual4repr}) representation these matrices appear in the
reduced density matrices \beq\label{redsur1} {\varrho}_{12}
=\mathcal{ L}\mathcal{ L}^{\dagger},\qquad
\overline{{\varrho}}_{34} =\mathcal{ L}^{\dagger}\mathcal{ L} \eeq
\noindent \beq \label{redsur2}{\varrho}_{13} =\mathcal{
M}\mathcal{ M}^{\dagger},\quad \overline{{\varrho}}_{24}
=\mathcal{ M}^{\dagger}\mathcal{ M} \eeq \noindent
\beq{{\varrho}}_{14} =\mathcal{ N}\mathcal{ N}^{\dagger},\qquad
\overline{{\varrho}}_{23} =\mathcal{ N}^{\dagger}\mathcal{
N}\label{redsur3}\eeq\noindent coming from the corresponding
operators like $\hat{\varrho}_{12}\equiv{\rm Tr}_{34}\vert
\psi\rangle\langle\psi\vert$.

We can embed these density matrices inside of a $28\times 28$ one
as follows. Write
$\vert\psi\rangle=\frac{1}{4!}Z_{ijkl}\hat{p}^{ijkl}\vert
0\rangle$ with only the relevant $16$ nonzero complex amplitudes
($ijkl\in \{1,2,\dots,\overline{4}\}$). Define the two-partite
reduced density matrix as \beq {\varrho^{ij}}_{kl}\equiv
\frac{1}{2}\langle\psi\vert\hat{p}^{ij}\hat{n}_{kl}\psi\rangle.
\label{2reddens} \eeq \noindent Explicitly one has \beq
{\varrho^{ij}}_{kl}=\frac{1}{4}\overline{Z}^{ijmn}Z_{lkmn}.
\label{hasznalni1} \eeq \noindent By virtue of this it can be
checked that if $\langle\psi\vert\psi\rangle=1$ then
${\varrho^{ij}}_{ij}=6$, i.e. ${\varrho^{ij}}_{kl}$ satisfies the
usual L\"owdin normalization adopted by quantum
chemists\cite{Lowdin}. From the $28$ independent index pairs $ij$
only $24$ gives nonzero contribution (pairs like
$1\overline{1},2\overline{2},3\overline{3},4\overline{4}$ give
zero), and similarly for the index pairs $kl$ one only has to take
into consideration  $24$ ones. It is easy to see that using for this
$24\times 24$ block the somewhat unusual labelling for the rows
and columns as $
(12,1\overline{2},\overline{1}2,\overline{1}\overline{2},\dots,
34,3\overline{4},\overline{3}4,\overline{3}\overline{4})$ we are
left with a block diagonal matrix consisting of six $4\times 4$
blocks. These are precisely the six reduced density matrices of
Eq.(\ref{redsur1})-(\ref{redsur3}). Since these are all having
trace equals to one, the trace of ${\varrho^{ij}}_{kl}=6$ as it
has to be.

Consider now one-half of the covariant ${\mathcal{R}^{ij}}_{kl}$
of Eq.(\ref{katamatrix}) i.e. \beq
\frac{1}{2}{\mathcal{R}^{ij}}_{kl}=\frac{1}{2}(\psi,\hat{p}^{ij}\hat{n}_{kl}\psi)
\label{katamasikalak} \eeq \noindent
 with explicit form given by
 Eq.(\ref{katanovasmatrix}) i.e. \beq \frac{1}{2}{\mathcal{R}^{
ij}}_{kl}=\frac{1}{4}\ast Z^{ijmn}Z_{lkmn}
\label{katanovasmatrixagain} \eeq \noindent where $\ast \psi$
denotes the Hodge dual of the four-form $\psi$.

\subsection{Majorana fermions}

Comparing Eqs.(\ref{2reddens}) and
(\ref{katamasikalak}) we see that our covariant and the
two-partite reduced density matrix is of the same structure up to
the important difference that the former features the bilinear
pairing and the latter the usual Hermitian scalar product. Now it
is known that this structural similarity is related to the
generalization\cite{SarLevlegujabb} of the usual Wootters spin
flip operation\cite{Wootters} as follows. For a spinor $\psi$
define its {\it spined flipped spinor} $\tilde{\psi}$ via the
formula \beq \langle\tilde{\psi}\vert\phi\rangle\equiv
(\psi,\phi). \label{spinflipdef}\eeq \noindent Then writing
$\vert\psi\rangle$ and $\vert\phi\rangle$ as in Eq.(\ref{elemek})
we obtain the result \beq \vert\tilde{\psi}\rangle
=\left(\hat{\Psi}^T\right)^{\dagger}\vert{\rm top}\rangle,
\label{explicitflip} \eeq \noindent where $\vert{\rm top}\rangle$
is defined in Eq.(\ref{TOP}) and clearly
$\hat{n}_i^{\dagger}=\hat{p}^i$. In our special case \beq
\vert\tilde{\psi}\rangle=
\frac{1}{4!}\tilde{Z}_{ijkl}\hat{p}^{ijkl}\vert 0\rangle=
\frac{1}{4!}\ast\overline{Z}_{ijkl}\hat{p}^{ijkl}\vert
0\rangle,\qquad
\tilde{Z}_{ijkl}=\ast\overline{Z}_{ijkl}=\frac{1}{4!}\varepsilon_{ijklabcd}\overline{Z}^{abcd}.
\label{flip4fermi} \eeq \noindent Now notice that the condition
\beq \vert\tilde{\psi}\rangle=\vert\psi\rangle\label{Majorana}
\eeq\noindent is a {\it reality condition} imposed on our
originally {\it complex} spinor. It can be checked that this
condition is precisely the {\it Majorana condition} for spinors\cite{Farrill}. In the following
a
spinor satisfying Eq.(\ref{Majorana}) will be referred to a {\it Majorana
spinor}. Hence we obtained the nice result that the generalized
Wootters spin flip operation of Ref.\cite{SarLevlegujabb} is
naturally related to the notion of Majorana spinors. As an extra
bonus of embedding qubits into Fock space we have managed to
understand the meaning of the well-known Wootters spin flip
operation of quantum information as a special case of an operation
related to a natural reality condition for spinors. In our special
case after comparing Eqs.(\ref{hasznalni1}) and
(\ref{katanovasmatrixagain}) we see that for a Majorana spinor
(Wootters self-conjugate spinor) our basic covariant
${\mathcal{R}^{ij}}_{kl}$ (the covariant of
Katanova\cite{Katanova}) is up to a factor of two just the
two-partite reduced density matrix ${\varrho^{ij}}_{kl}$.

Let us see how these findings manifest themselves in the special
case of embedded four-qubits. Now the independent components of
${\mathcal{R}^{ij}}_{lk}$ form a $28\times 28$ matrix\footnote{Notice that the matrix $\mathcal{R}$ is defined with a swap of the indices $kl$ hence a sign change.} ${\mathcal{R}}$ whose
structure is similar to the one of ${\varrho^{ij}}_{lk}$. Namely
this matrix is block diagonal and consists of seven $4\times 4$
blocks. One block is containing merely zeros, the remaining six
ones are  \beq
\mathcal{R}_{12}\equiv\epsilon\otimes\epsilon\mathcal{L}\epsilon\otimes\epsilon
\mathcal{L}^t,\qquad
\mathcal{R}_{34}\equiv\epsilon\otimes\epsilon\mathcal{L}^t\epsilon\otimes\epsilon
\mathcal{L} \label{12r} \eeq \noindent \beq
\mathcal{R}_{13}\equiv\epsilon\otimes\epsilon\mathcal{M}\epsilon\otimes\epsilon
\mathcal{M}^t,\quad
\mathcal{R}_{24}\equiv\epsilon\otimes\epsilon\mathcal{M}^t\epsilon\otimes\epsilon
\mathcal{M} \label{23r} \eeq \noindent \beq
\mathcal{R}_{14}\equiv\epsilon\otimes\epsilon\mathcal{N}\epsilon\otimes\epsilon
\mathcal{N}^t,\quad
\mathcal{R}_{23}\equiv\epsilon\otimes\epsilon\mathcal{N}^t\epsilon\otimes\epsilon
\mathcal{N} \label{13r} \eeq \noindent where \beq
\epsilon\otimes\epsilon=\begin{pmatrix}0&1\\-1&0\end{pmatrix}\otimes
\begin{pmatrix}0&1\\-1&0\end{pmatrix}=
\begin{pmatrix}0&0&0&1\\0&0&-1&0\\0&-1&0&0\\1&0&0&0\end{pmatrix}
 \label{innermatr}\eeq \noindent and $\mathcal{L}^t$ refers to the transposed
matrix of $\mathcal{L}$ (not to be confused with the transposition
used for spinors in expressions like $\hat{\Psi}^T$ in
(\ref{explicitflip}) see also Eq.(\ref{transposed})).
Explicitly we have
\beq
\mathcal{R}=\frac{1}{2}\begin{pmatrix}\mathcal{R}_{12}& \ & \ & \ & \ & \ & \ \\
\ & \mathcal{R}_{34} & \ & \ & \ & \ & \ \\
\ & \ & \mathcal{R}_{13} & \ & \ & \ & \ \\
\ & \ & \ & \mathcal{R}_{24} & \ & \ & \ \\
\ & \ & \ & \ & \mathcal{R}_{14} & \ & \  \\
\ & \ & \ & \ & \ & \mathcal{R}_{23} & \  \\
\ & \ & \ & \ & \ & \ & 0\\
\end{pmatrix}.
\label{blokkalak}
\eeq
\noindent
Now for a $4$
qubit state of the form (\ref{usual4repr}) the amplitudes of the
usual Wootters spin flipped state is defined as \beq
\tilde{\psi}_{\mu_1\mu_2\mu_3\mu_4}=
\epsilon^{\mu_1\mu^{\prime}_1}\epsilon^{\mu_2\mu^{\prime}_2}\epsilon^{\mu_3\mu^{\prime}_3}
\epsilon^{\mu_4\mu^{\prime}_4} \overline{\psi}_{\mu^{\prime}_1
\mu^{\prime}_2\mu^{\prime}_3\mu^{\prime}_4}. \label{Wottusual}
\eeq \noindent After representing the $16$ amplitudes in terms
of $4\times 4$ matrices the Wotters spin flip corresponds to
transformations like
$\mathcal{L}\mapsto\epsilon\otimes\epsilon\overline{\mathcal{L}}\epsilon\otimes\epsilon$
terms showing up in Eqs.(\ref{12r})-(\ref{13r}).

\subsection{Recovering the basic four-qubit invariants}

Let us now try to recover the basic four-qubit invariants in our
spinorial framework. From the work of Luque and Thibon\cite{Luque}
we know that for four qubits we have four algebraically
independent invariants. One of them is of order 2 ($H$) two of
order 4 ($L$ and $M$) and one of order six ($D$). For a pair of
arbitrary vectors $U,V\in\mathbb{C}^4$ let us define an inner
product by the formula \beq U\cdot
V=U^0V^3-U^1V^2-U^2V^1+U^3V^0\label{innerpr} \eeq \noindent i.e.
the inner product has the matrix of Eq.(\ref{innermatr}). Then in
the notation of Eq.(\ref{Lmatrix4qbit}) we have the formula for
$H$ \beq H=U\cdot Z-V\cdot
W=\epsilon^{\mu_1\mu^{\prime}_1}\epsilon^{\mu_2\mu^{\prime}_2}\epsilon^{\mu_3\mu^{\prime}_3}
\epsilon^{\mu_4\mu^{\prime}_4}\psi_{\mu_1 \mu_2\mu_3\mu_4}
\psi_{\mu^{\prime}_1
\mu^{\prime}_2\mu^{\prime}_3\mu^{\prime}_4}=\frac{1}{2}(\psi,\psi)
. \label{explH} \eeq \noindent Notice that the first expression
for $H$ is hiding its permutation invariance, clearly displayed by
the second. The third expression relates this important invariant
to our bilinear form defined for spinors see Eq.(\ref{invkvad4}),
where in this case for the labelling of the amplitudes
Eq.(\ref{ezitta4qbit}) should be used.

For the definitions of the fourth order invariants we have \beq
L={\rm Det}\mathcal{L},\quad M={\rm Det}\mathcal{M},\quad N={\rm
Det}\mathcal{N}. \label{Thiboninv} \eeq \noindent and the
important identity\cite{Luque} \beq L+M+N=0.\label{nemfugtlen}
\eeq \noindent

In order to define the sixth order invariant $D$ we use the characteristic polynomial of the matrix $R\equiv \mathcal{R}_{12}$ as a generating polynomial for the algebraically independent invariants\cite{Levfour}
\beq
{\mathcal{P}}(R,t)\equiv{\rm Det}(tI-R)=t^4-s_1t^3+s_2t^2-s_3t +s_4
\eeq
\noindent
where
\beq
s_1={\rm Tr}R=2H,
\label{newton1}
\eeq
\noindent
\beq
2s_2=({\rm Tr}R)^2-{\rm Tr}R^2 = H^2+4M+2L
\label{newton2}
\eeq
\noindent
\beq
 3!s_3=
 ({\rm Tr}R)^3-3{\rm Tr}R{\rm Tr}R^2+2{\rm Tr}R^3=
 4D+2HL,
\label{newton3}
\eeq
\noindent
\beq
4!s_4=({\rm Tr}R)^4+8{\rm Tr}R{\rm Tr}R^3+3({\rm Tr}R^2)^2-6({\rm Tr}R)^2{\rm Tr}R^2-6{\rm Tr}R^4=4!{\rm Det}R=4!L^2.
\label{newton4}
\eeq
\noindent
An explicit computation shows that\cite{Levfour}
\beq
s_3=2{\rm Det}\begin{pmatrix}U\cdot U&U\cdot V&U\cdot Z\\
U\cdot W&V\cdot W&W\cdot Z\\
U\cdot Z&V\cdot Z&Z\cdot Z\end{pmatrix}-
2{\rm Det}\begin{pmatrix}U\cdot V&V\cdot V&V\cdot W\\
U\cdot W&V\cdot W&W\cdot W\\
U\cdot Z&V\cdot Z&W\cdot Z\end{pmatrix} \label{sexkiirva} \eeq
\noindent
which imlicitly defines $D$.
For the algebraically independent set of $SL(2,\mathbb{C})^{\otimes 4}$ invariants either the set $s_1,s_2,s_3,s_4$ or the one $H,L,M,D$ can be used.
There is yet another way of looking at the sixth order invariants which will be useful. Using $H,L,M$ and $D$ one can define new sixth order combinations\cite{Luque}
$E$ and $F$ as follows
\beq
D=E-HL,\qquad E=F-HN,\qquad F=D-HM.
\label{kombisex}
\eeq
\noindent
Clearly these combinations are related by permutation symmetry of the qubits
in a cyclic manner.
This has the important corollary that if in the characteristic polynomial we plug in for $R$ either of the matrices in Eq.(\ref{12r})-(\ref{13r}) based on $\mathcal{L},\mathcal{N}$ or $\mathcal{M}$
then the invariants $s_1,s_2,s_3,s_4$ showing up will always have a similar form with the letters
$L,M,N$ and $D,E,F$ cyclically permuted.
Explicitly
\beq
s_1=\left\{\begin{array}{rcl} 2H &\mbox{using} & {\mathcal{ L}}\\
2H,&\mbox{using} & {\mathcal{ N}}\\2H,
&\mbox{using} & {\mathcal{ M}} \end{array}\right.
\label{esek1}
\eeq
\noindent
\beq
s_2=\left\{\begin{array}{rcl} H^2+2(M-N) &\mbox{using} & {\mathcal{ L}}\\
H^2+2(L-M),&\mbox{using} & {\mathcal{ N}}\\H^2+2(N-L),
&\mbox{using} & {\mathcal{ M}} \end{array}\right.
\label{esek2}
\eeq
\noindent
\beq
s_3=\left\{\begin{array}{rcl} 2(D+E) &\mbox{using} & {\mathcal{ L}}\\
2(E+F),&\mbox{using} & {\mathcal{ N}}\\2(F+D),
&\mbox{using} & {\mathcal{ M}} \end{array}\right.
\label{esek3}
\eeq
\noindent
\beq
s_4=\left\{\begin{array}{rcl} L^2 &\mbox{using} & {\mathcal{ L}}\\
N^2,&\mbox{using} & {\mathcal{ N}}\\M^2, &\mbox{using} &
{\mathcal{ M}} \end{array}\right. \label{esek4} \eeq \noindent
where for arriving at this form with permutation symmetry
displayed we used the identities (\ref{nemfugtlen}) and
(\ref{kombisex}). Recall now that from the terms showing up in
these expressions one can form four algebraically independent
combinations which apart from $SL(2,\mathbb{C})^{\times 4}$
invariance displaying permutation invariance as well. These form
the set\cite{Schlafli,Luque} $\{H,\Sigma,\Gamma,\Pi\}$ where \beq
\Sigma= L^2+M^2+N^2,\qquad \Gamma = D+E+F,\qquad \Pi=
(L-M)(M-N)(N-L).\label{schlaflipermu} \eeq \noindent

In order to reveal the spinorial origin of these
invariants one calculates the traces of the relevant $4\times 4$
blocks of the basic covariant ${\mathcal{R}^{ij}}_{kl}$ associated to the $28\times 28$ matrix $\mathcal{R}$ of Eq.(\ref{blokkalak}) with explicit structure given by
Eqs.(\ref{12r})-(\ref{13r}).

\begin{subequations}
\label{resek}
\beq
\frac{1}{2}{\rm Tr}R=
\left\{\begin{array}{rcl} H &\mbox{using} & {\mathcal{ L}}\\
H,&\mbox{using} & {\mathcal{ N}}\\H,
&\mbox{using} & {\mathcal{ M}} \end{array}\right.
\label{resek1}
\eeq
\noindent
\beq
\frac{1}{2}{\rm Tr}R^2=
\left\{\begin{array}{rcl} H^2+2(N-M) &\mbox{using} & {\mathcal{ L}}\\
H^2+2(M-L),&\mbox{using} & {\mathcal{ N}}\\H^2+2(L-N),
&\mbox{using} & {\mathcal{ M}} \end{array}\right.
\label{resek2}
\eeq
\noindent
\beq
\frac{1}{2}{\rm Tr}R^3
=\left\{\begin{array}{rcl} H^3+6H(N-M)+3(D+E) &\mbox{using} & {\mathcal{ L}}\\
H^3+6H(M-L)+3(E+F)
,&\mbox{using} & {\mathcal{ N}}\\
H^3+6H(L-N)+3(F+D)
,
&\mbox{using} & {\mathcal{ M}} \end{array}\right.
\label{resek3}
\eeq
\beq
\frac{1}{2}{\rm Tr}R^4
=\left\{\begin{array}{rcl} H^4+12H^2(N-M)+
8H(D+E)+
4(N-M)^2-2L^2
&\mbox{using} & {\mathcal{ L}}\\
H^4+
12H^2(M-L)+8H(E+F)+4(M-L)^2-2N^2
,&\mbox{using} & {\mathcal{ N}}\\
H^4+12H^2(L-N)+8H(F+D)+4(L-N)^2-2M^2
.
&\mbox{using} & {\mathcal{ M}} \end{array}\right.
\label{resek4}
\eeq
Further traces of powers can be calculated, here we merely give the expressions for the fifth and sixth powers we need later.
\begin{eqnarray}
\frac{1}{2}{\rm Tr}R^5&=&H^5+
20H^3(N-M)+
15H^2(D+E)\nonumber\\&-&5HL^2+
20H(N-M)^2+
10(D+E)(N-M)
\label{otodtr}
\end{eqnarray}
\noindent
\begin{eqnarray}
\frac{1}{2}{\rm Tr}R^6&=&H^6+30H^4(N-M)
+24H^3(D+E)-9H^2L^2+60H^2(N-M)^2\nonumber\\
&+&48H(D+E)(N-M)+6(E+D)^2+8(N-M)^2-6L^2(N-M)
\end{eqnarray}
\noindent
where in these expressions we have merely displayed the $R=\mathcal{R}_{12}$
choice.
\end{subequations}

Let us new define  $SL(2,\mathbb{C})^{\times 4}\rtimes S_4$ invariants
$g_{2p}, p=1,2,\dots$ as follows
\beq
g_{2p}=\frac{1}{2}\sum_{a<b}{\rm Tr}\mathcal{R}_{ab}^p,
\label{geinvar}
\eeq
\noindent
i.e. in order to form these invariants we have to add the traces of the powers
of the six nontrivial matrices showing up in Eq.(\ref{blokkalak}).
Then a straightforward calculation shows that
\beq
g_2=6H,\qquad g_4=6H^2,\qquad g_6=6H^3+12\Gamma
\label{246}
\eeq
\noindent
\beq
g_8=6H^4+32H\Gamma+20\Sigma
\label{8}
\eeq
\noindent
\beq
g_{10}=6H^5+90H\Sigma+60H^2\Gamma
\label{10}
\eeq
\noindent
\beq
g_{12}=6H^6+96H^3\Gamma+250H^2\Sigma+16\Gamma^2-60\Pi.
\label{12}
\eeq
\noindent
From this it follows that as an independent set of generators the set
\beq
\{g_2,g_6,g_8,g_{12}\}
\label{inepgs}
\eeq
\noindent
can be used.
In order to relate our set of generators to the one of Chen et.al.\cite{Chen}
we express the invariant $g_{10}$ in terms of the independent ones. The result is
\beq
2^5\cdot 3^4g_{10}=7g_2^5+2^3\cdot 3^5 g_2g_8-2^3\cdot 7\cdot 9 g_2^2g_6.
\label{levaydjoko}
\eeq
\noindent
Now comparing this equation with Eq.(11) of Ref.\cite{Chen} one concludes that
the set of independent generators used by Chen et.al. namely $\{f_2^{\prime},
f_6^{\prime},f_8^{\prime},f_{12}^{\prime}\}$ is related to ours simply
\beq
f_{2p}^{\prime}=2^{1-p}g_{2p}={\rm Tr}\mathcal{R}^p
\label{djokolevay}
\eeq
\noindent
where $\mathcal{R}$ is the matrix of Eq.(\ref{blokkalak}).
Now the invariants $I_{2p}$ of Eq.(\ref{fermiinvi})
are just trivial multiples of $f_{2p}^{\prime}$ namely
\beq
I_{2p}=(-1)^p2^pf_{2p}^{\prime},
\label{sl8inviembed}
\eeq
\noindent
and the fermionic invariants $J_{2p}$ can also be calculated
using Eq.(\ref{Katanovavegso}).
Clearly $J_{2p}$ will be again a polynomial of the set $\{f_2^{\prime},
f_6^{\prime},f_8^{\prime},f_{12}^{\prime}\}$ which we will not give here.

\section{The algebra of ${\rm Spin}(16,\mathbb{C})$ invariant polynomial functions}

Let $\mathcal{A}$ be the algebra of complex polynomial functions on either $\mathcal{F}_+$ or $\mathcal{F}_-$ i.e. on the $128$ dimensional complex vector space space of Weyl spinors of definite chirality which are invariant under $G_0={\rm Spin}(16,\mathbb{C})$.
One can define the affine variety $\mathcal{F}_+/{\rm Spin}(16,\mathbb{C})$ associated to $\mathcal{A}$.
One can then show that this variety is isomorphic\cite{Antonyan2} to $\mathbb{C}^8$.
In this section we would like to elaborate on the structure of the $8$ algebraically independent generators
for $\mathcal{A}$.
From the physical point of view the magnitudes of these generators will give possible measures of entanglement, that are invariant under the generalized SLOCC subgroup $G_0$.

Note that such an investigation can be 
regarded as a natural generalization of 
the one initiated in Ref.\cite{Chen} where 
four fermions with $8$ modes were considered. 
In this case the corresponding algebra 
of invariants $\mathcal{B}$ is the one of complex 
polynomial functions on the $70$ dimensional complex 
vector space $\wedge^4V^{\ast}$ with $V=\mathbb{C}^8$ invariant 
under $SL(8,\mathbb{C})$. The latter group is the nontrivial subgroup of the SLOCC group and  the affine
variety $\wedge^4V^{\ast}/SL(8,\mathbb{C})$ is isomorphic to the affine space
$\mathbb{C}^7$. In our fermionic formalism the seven generators of $\mathcal{B}$ are of the form\cite{Katanova,Chen} $f_{2p}={\rm Tr}\mathcal{R}^p$ with $p=1,3,4,5,6,7,9$ where $\mathcal{R}$ is a $28\times 28$ matrix not subject to the restrictions displayed in Eq.(\ref{blokkalak}).
Furthermore, for embedded four-qubit systems our detailed calculations based on the special form of Eq.(\ref{blokkalak}) show how  the algebra $\mathcal{C}$ of complex polynomial functions on $\mathbb{C}^2\otimes
\mathbb{C}^2\otimes
\mathbb{C}^2\otimes
\mathbb{C}^2$ invariant under $SL(2,\mathbb{C})^{\times 4}\rtimes S_4$
are derived from the basic femionic invariants.
In this case the corresponding affine variety is isomorphic to $\mathbb{C}^4$
with generators
$\{g_2,g_6,g_8,g_{12}\}$ of Eq.(\ref{246})-(\ref{12}).
According to Eq.(\ref{djokolevay}) these generators also correspond to the
set\cite{Chen}
$\{f_2^{\prime},
f_6^{\prime},f_8^{\prime},f_{12}^{\prime}\}$.
Clearly we have a sequence of embedded algebras
\beq
\mathcal{C}\subset\mathcal{B}\subset\mathcal{A}.
\label{algembed}
\eeq
\noindent
In Ref.\cite{Chen} the restriction map between the algebras $\mathcal{C}\subset\mathcal{B}$ has been studied.
An aim of this section is to initiate a study concerning the algebra $\mathcal{A}$, as an object naturally incorporating all cases.

Now the results for the embedding $\mathcal{C}\subset\mathcal{B}$ follow
from the decompositions based the symmetric spaces
\beq
\mathfrak{so}_8=(\mathfrak{sl}_2\oplus
\mathfrak{sl}_2\oplus
\mathfrak{sl}_2\oplus
\mathfrak{sl}_2)
\oplus\mathfrak{m}_{16},
\label{rom1}
\eeq
\noindent
\beq
\mathfrak{e}_7=\mathfrak{sl}_8\oplus\mathfrak{m}_{70}.
\label{rom2}
\eeq
\noindent
Note that results for the
symmetric space decomposition
\beq
\mathfrak{e}_8=\mathfrak{so}_{16}\oplus\mathfrak{m}_{128}
\label{rom3}
\eeq
\noindent
are also available in the literature\cite{Antonyan2}.
Since the subspace $\mathfrak{m}_{128}$ in this approach is just the space of Weyl spinors i.e. $\mathcal{F}_+$ this observation
enables an explicit exploration of the structure of the restriction map for the generators of the algebra $\mathcal{A}$ an issue which is the subject of the next subsections.

\subsection{The semisimple orbit for four qubits}

In order to gain some insight into the structure of $\mathcal{A}$ we
reformulate some results already discussed in the literature.
Take the representative of the semisimple orbit of four qubit states in the form\cite{Gour1}
\beq
\vert \mathcal{G}(x)\rangle\equiv \sum_{\alpha=1}^4x_{\alpha}\vert\phi_{\alpha}\rangle,\qquad
\vert\phi_{\alpha}\rangle\equiv \vert\varphi_{\alpha}\rangle\otimes\vert\varphi_{\alpha}\rangle,\qquad x\equiv (x_1,x_2,x_3,x_4)\in \mathbb{C}^4
\label{semisimple}
\eeq
\noindent
where
\beq
\vert\varphi_1\rangle=\frac{1}{\sqrt{2}}\left(\vert 00\rangle +\vert 11\rangle
\right)
\qquad\vert\varphi_2\rangle=\frac{1}{\sqrt{2}}\left(\vert 01\rangle -\vert 10\rangle\right)
\label{Bellbasis1}
\eeq
\noindent
\beq
\vert\varphi_3\rangle=\frac{1}{\sqrt{2}}\left(\vert 01\rangle +\vert 10\rangle
\right)
\qquad\vert\varphi_4\rangle=\frac{1}{\sqrt{2}}\left(\vert 00\rangle -\vert 11\rangle\right).
\label{Bellbasis2}                                                              \eeq                                                                            \noindent
Alternatively one can write
\beq
\vert \mathcal{G}(x)\rangle=y_1(\vert 0000\rangle +\vert 1111\rangle)+
                              y_2(\vert 0011\rangle +\vert 1100\rangle)+
                              y_3(\vert 0101\rangle +\vert 1010\rangle)+
                              y_4(\vert 0110\rangle +\vert 1001\rangle),
\label{semialtern}
\eeq
\noindent
where
\beq x_1=y_1+y_4,\qquad x_2=y_3-y_2,\qquad x_3=y_3+y_2,\qquad x_4=y_1-y_4.
\label{ipszilonok}
\eeq
\noindent
Take the following $24$ element set of elementary polynomials in $x$
\beq
\pm 2x_1,\quad\pm2x_2,\quad\pm2x_3,\quad\pm 2x_4,\quad\pm x_1\pm x_2\pm x_3\pm x_4
\label{24roots}
\eeq
\noindent
where the last item in the list refers to all of the $16$ possible sign combinations.
Call these $24$ elementary polynomials $e_s(x),\quad s=1,2,\dots 24$.
Let us define the new polynomials
\beq
\pi_{2p}(x)\equiv \sum_{s=1}^{24}[e_s(x)]^{2p}
\label{pipol}
\eeq
\noindent
Then for example
one has
\beq
\pi_2(x)=24\sum_{\alpha=1}^4x_{\alpha}^2
\label{pi2}
\eeq
\noindent
\beq
\pi_{6}(x)=48\left[3\sum_{\alpha =1}^4x_{\alpha}^6 +5\sum_{\alpha\neq\beta}x_{\alpha}^2x_{\beta}^4+
30\sum_{\alpha>\beta>\gamma}x_{\alpha}^2x_{\beta}^2x_{\gamma}^2\right].
\label{pi6}
\eeq
\noindent
Now a calculation of the simplest two invariants $H$ and $\Gamma$ for the state
Eq.(\ref{semisimple}) shows that\cite{Djok}
\beq
2H=\sum_{\alpha =1}^4x_{\alpha}^2,\qquad
2^5\Gamma=\sum_{\alpha =1}^4x_{\alpha}^6-\sum_{\alpha\neq\beta}x_{\alpha}^2x_{\beta}^4+18\sum_{\alpha>\beta>\gamma}x_{\alpha}^2x_{\beta}^2x_{\gamma}^2
\label{Gamsemi}
\eeq
\noindent

Comparing now $\pi_{2p}$ and $g_{2p}$ of Eq.(\ref{246}) for $p=1,3$ one gets that
\beq
\pi_2(x)=2^3g_2(x),\qquad \pi_{6}(x)=2^7g_6(x).
\label{conjectu}
\eeq
\noindent
A computer check shows that this simple pattern survives hence
\beq
\pi_{2p}(x)=2^{2p+1}g_{2p}(x),\qquad p=1,3,4,6.
\label{pattern1}
\eeq
\noindent
or alternatively
\beq
\pi_{2p}(x)=(-1)^p2^{2p}I_{2p}
\label{fontosconj}
\eeq
\noindent
where $I_{2p}$ are the fermionic invariants defined in Eq.(\ref{fermiinvi})
.

Notice that using the new parametrization of Eq.(\ref{ipszilonok})
for the $e_r$ we immediately get the simple expressions \beq
g_{2p}=\sum_{\alpha<\beta}(y_{\alpha}+y_{\beta})^{2p}+
\sum_{\alpha<\beta}(y_{\alpha}-y_{\beta})^{2p}, \qquad
\alpha,\beta=1,2,3,4. \label{Wallachform} \eeq \noindent Notice
that apart from a factor of $\frac{1}{6}$ and a different
labelling convention used this expression for $g_{2p}$ is of the
same form as the invariants $\mathcal{F}_{2p}$ of  Eq.(1) of Gour
and Wallach\cite{Gour2}. It is also important to realize that
these polynomials for $p=1,3,4,6$ constitute a set of
algebraically independent polynomials\cite{Lee,Mehta} invariant
under the Weyl group of the exceptional group $F_4$. This
interesting connection between the dense orbit of four qubit
states and $F_4$ was emphasized in Ref.\cite{Gour1,Gour2}.

However, the  algebraically independent
sets $\{\mathcal{F}_2,\mathcal{F}_6,\mathcal{F}_8,\mathcal{F}_{12}\}$
and  $\{g_2,g_6,g_8,g_{12}\}$ are not the same.
Indeed in Ref.\cite{Gour2} instead of our parameters $x_{\alpha}$  the parametrization
\beq
(z_0,z_1,z_2,z_3)=(x_1,x_4,x_3,x_2)
\label{corrwall}
\eeq
\noindent
was used.
In this parametrization\cite{Gour2}
\beq
\mathcal{F}_{2p}=\frac{1}{6}\sum_{\alpha<\beta}(z_{\alpha}-z_{\beta})^{2p}-
\frac{1}{6}\sum_{\alpha<\beta}(z_{\alpha}-z_{\beta})^{2p},\qquad \alpha,\beta=0,1,2,3.
\label{efwallach}
\eeq
\noindent
Hence according to Eq.(\ref{corrwall}) our polynomials $g_{2p}$ expressed in terms of the parameters $y_{\alpha},\alpha=1,2,3,4$ are of the same form as the polynomials $\mathcal{F}_{2p}$ expressed in terms of the parameters $x_{\alpha}, \alpha=1,2,3,4$.

As a result the explicit forms of the polynomials $\mathcal{F}_{2p}$ expressed in terms of the set $\{H,\Gamma,\Sigma,\Pi\}$ should be some different combinations then the ones shown in Eqs.(\ref{246})-(\ref{12}).
For example a quick calculation shows that
$\mathcal{F}_{6}=12H^3-16\Gamma$  on the other hand according to Eq.(\ref{246})
$g_6=6H^3+12\Gamma$.
This is in accord with the result found Section IV. of Ref.\cite{Djok2}
\beq
\mathcal{F}_2=2H,\qquad\mathcal{F}_{6}=4(3H^3-4\Gamma),
\label{f26}
\eeq
\noindent
\beq
\mathcal{F}_8=\frac{4}{3}(33H^4-104H\Gamma+40\Sigma)
\label{f8}
\eeq
\noindent
\beq
\mathcal{F}_{12}=\frac{4}{3}(513H^6-3012H^3\Gamma+2180H^2\Sigma+488\Gamma^2+480\Pi).
\label{f12}
\eeq
\noindent
Comparing these expressions with the ones of Eqs.(\ref{246})-(\ref{12}) we see
that our generating set $\{g_2,g_6,g_8,g_{12}\}$ is more elegant as the expansion coefficients are much simpler and they can be seen as the ones derived from a more general procedure based on fermionic systems as spinors.
Notice however, that according to  Eq.(\ref{djokolevay}) up to $2^{1-p}$ this generator system is the same as the one $\{f_2^{\prime},f_{6}^{\prime},f_8^{\prime},f_{12}^{\prime}\}$ which already appeared in Ref.\cite{Chen} as the one coming from the invariants of Katanova\cite{Katanova}.
Here we added to these results a further twist by also displaying their explicit form in terms of the usual set $\{H,\Gamma,\Sigma,\Pi\}$ originally due to Schl\"afli\cite{Schlafli}.
We also note that apart from a factor  of $6$ the set $\{\mathcal{F}_2,\mathcal{F}_6,\mathcal{F}_8,\mathcal{F}_{12}\}$
is the same as the Saito-Sekiguchi set\cite{Sato} of generators a point emphasized
 in the Appendix of Ref.\cite{Djok}.

\subsection{Representing the dense orbit under  $\mathbb{C}^{\times}\times{\rm Spin}(16,\mathbb{C})$}

Our aim here is to use an eight parameter representative of the generic orbit in $\mathcal{F}_+$
under the action of the generalized SLOCC group $\mathbb{C}^{\times}\times{\rm Spin}(16,\mathbb{C})$ for obtaining explicit forms for the invariants.
In Ref.\cite{Chen} it was shown how the four parameter family of states of Eq.(\ref{semisimple}) can be embedded into a seven parameter family belonging to $\wedge^4\mathbb{C}^8$. This means that this family can be regarded as restrictions of a more general one  for four fermions with eight modes.
This chain of generalizations is based on the (\ref{rom1})-(\ref{rom3}) sequence of Lie algebras.
Here we give a spinorial entanglement based generalization of the $E_8$ case. Note that our representative of the relevant entanglement class is equivalent to the representative of the standard semisimple orbit already known in the mathematics literature\cite{Antonyan2}.

Let us consider the Fock space version of the state of Eq.(\ref{semialtern})
\begin{eqnarray}
\vert \mathcal{G}\rangle &=&y_1(\hat{p}^{1234}+\hat{p}^{\overline{1234}})+
y_2(\hat{p}^{12\overline{34}}+\hat{p}^{\overline{12}34})\nonumber\\
&+&y_3(\hat{p}^{1\overline{2}3\overline{4}}+\hat{p}^{\overline{1}2\overline{3}4})
+y_4(\hat{p}^{1\overline{23}4}+\hat{p}^{\overline{1}23\overline{4}})\vert 0\rangle=\sum_{\alpha=1}^4y_{\alpha}\vert E_{\alpha}\rangle
\label{4paramversion}
\end{eqnarray}
\noindent
where
\beq
\vert E_1\rangle=(\hat{p}^{1234}+\hat{p}^{\overline{1234}})\vert 0\rangle,\qquad
\vert E_2\rangle=
(\hat{p}^{12\overline{34}}+\hat{p}^{\overline{12}34})\vert 0\rangle
\label{Ek1}
\eeq
\noindent
\beq
\vert E_3\rangle=(\hat{p}^{1\overline{2}3\overline{4}}+\hat{p}^{\overline{1}2\overline{34}})\vert 0\rangle,\qquad
\vert E_4\rangle=
(\hat{p}^{1\overline{23}4}+\hat{p}^{\overline{1}23\overline{4}})\vert 0\rangle.
\label{Ek2}
\eeq
\noindent

Now if we make the identification
\beq
\{1,2,3,4,\overline{1},\overline{2},\overline{3},\overline{4}\}\equiv\{13572468\}
\label{identdjoki}
\eeq
\noindent
then the basis vectors $p_2,p_4,p_5$ and $-p_6$ of Ref.\cite{Chen,Antonyan1} will correspond to the ones $\vert E_{\alpha}\rangle$.
If according to (\ref{rom2}) we identify these states as four generators belonging to the $\mathfrak{m}_{70}$ part of the Lie algebra $\mathfrak{e}_7$ we see that they define four from the seven of the basis states of a seven dimensional Cartan subspace
$\mathfrak{c}$. In this notation the remaining three basis vectors (denoted by $p_1,p_3$ and $-p_7$ in Ref.\cite{Chen,Antonyan1}) have the form
\beq
\vert E_5\rangle=(\hat{p}^{1\overline{1}4\overline{4}}+\hat{p}^{2\overline{2}3\overline{3}})\vert 0\rangle,\qquad
\vert E_6\rangle=(\hat{p}^{1\overline{1}3\overline{3}}+\hat{p}^{2\overline{2}4\overline{4}})
\vert
 0\rangle,\qquad \vert E_7\rangle= (\hat{p}^{1\overline{1}2\overline{2}}+\hat{p}^{3\overline{3}4\overline{4}})\vert
  0\rangle.
\label{kilogo}
\eeq
\noindent
Notice now that in the notation of section 4.4. the basis vectors of the first kind ($p_2,p_4,p_5,-p_6$) are spanning a subspace of the {\it single occupancy subspace} $\mathcal{F}^{0000}\subset \mathcal{F}_+$. On the other
hand the ones of the second kind ($p_1,p_3,-p_7$) are spanning a subspace of the {\it double occupancy subspace} $\mathcal{F}^{1111}\subset\mathcal{F}_+$.
According to Eq.(\ref{8drb4q}) these subspaces are related by the action of the operator
\beq
\hat{\Omega}\equiv \hat{\Gamma}_1
\hat{\Gamma}_2\hat{\Gamma}_3\hat{\Gamma}_4
.
\label{Omegaoper}
\eeq
\noindent
Under the action of $\hat{\Omega}$ the basis vectors $\vert E_j\rangle, j=1,\dots 8$ are mapped to each other as
\beq
\vert E_{\alpha}\rangle \mapsto \vert E_{9-\alpha}\rangle ,\qquad\alpha=1,2,3,4
\label{ittjina8}
\eeq
\noindent
where we have introduced a new basis state $\vert E_8\rangle$
which is of the form
\beq
\vert E_8\rangle =(\hat{\bf{1}}+\hat{p}^{1234\overline{1234}})\vert 0\rangle.
\label{pe8}
\eeq
\noindent
Hence the double occupancy version of the state of Eq.(\ref{4paramversion}) is
\begin{eqnarray}
\vert\mathcal{G}^{\prime}\rangle&=&
y_5(\hat{p}^{1\overline{1}4\overline{4}}+\hat{p}^{2\overline{2}3\overline{3}}
\vert
0\rangle+
y_6
(\hat{p}^{1\overline{1}3\overline{3}}+\hat{p}^{2\overline{2}4\overline{4}})     \vert
 0\rangle\nonumber\\
&+&y_7
(\hat{p}^{1\overline{1}2\overline{2}}+\hat{p}^{3\overline{3}4\overline{4}})\vert 0\rangle
+y_8
(\hat{\bf{1}}+\hat{p}^{1234\overline{1234}})\vert 0\rangle
=\sum_{\alpha=1}^4y_{9-\alpha}\hat{\Omega}\vert E_{\alpha}\rangle.
\label{4paravesszo}
\end{eqnarray}
\noindent
Now the $8$ parameter family of states we would like to propose is of the form
\beq
\vert G(y)\rangle \equiv \sum_{\alpha=1}^4\left(y_{\alpha}+y_{9-\alpha}\hat{\Omega}\right)\vert E_{\alpha}\rangle.
\label{Klasszstate}
\eeq
\noindent
An alternative form of this state is
\beq
\vert G(y)\rangle=\left(y_8\hat{\bf{1}}+\frac{1}{4!}Z_{ijkl}\hat{p}^{ijkl}+
y_8\hat{p}^{1234\overline{1234}}\right)\vert 0\rangle
\label{alternatesemisimple}
\eeq
\noindent
where
\beq
Z_{1234}=Z_{\overline{1234}}=y_1,\qquad
Z_{12\overline{34}}=Z_{\overline{12}34}=y_2,
\label{z12}
\eeq
\noindent
\beq
Z_{1\overline{2}3\overline{4}}=Z_{\overline{1}2\overline{3}4}=y_3,\qquad
Z_{1\overline{23}4}=Z_{\overline{1}23\overline{4}}=y_4,
\label{z34}
\eeq
\noindent
\beq
Z_{1\overline{1}4\overline{4}}=Z_{2\overline{2}3\overline{3}}=y_5,\qquad
Z_{1\overline{1}3\overline{3}}=Z_{2\overline{2}4\overline{4}}=y_6,\qquad
Z_{1\overline{1}2\overline{2}}=Z_{3\overline{3}4\overline{4}}=y_7.\qquad
\label{z567}
\eeq
\noindent
Notice that if we define the {\it dual tensor} $\ast Z^{ijkl}$ as
\beq
\ast Z^{ijkl}\equiv\frac{1}{4!}\varepsilon^{ijklabcd}Z_{abcd}
\label{dual4form}
\eeq
\noindent
then we have
\beq
\ast Z^{1234}=Z_{\overline{1234}},\qquad
\ast Z^{\overline{1234}}=Z_{1234},\qquad
\ast Z^{12\overline{34}}=Z_{\overline{12}34},\qquad \dots\qquad \ast Z^{1\overline{1}2\overline{2}}=Z_{3\overline{3}4\overline{4}}
\label{dualexplic}
\eeq
\noindent
hence the tensor $Z_{ijkl}$ is {\it self-dual}.

We will need the matrix elements \beq
(G(y),G(y))=2y_8^2+\frac{1}{4!}\ast
Z^{ijkl}Z_{ijkl}=2\sum_{n=1}^8y_{n}^2 \label{invar} \eeq \noindent
\beq (G(y),\hat{p}^{ijkl}G(y))=2y_8\ast Z^{ijkl},\qquad
(G(y),\hat{n}_{ijkl}G(y))=2y_8 Z_{lkji}. \label{kovar2} \eeq
\noindent \beq (G(y),\hat{p}^{ij}\hat{n}_{kl}G(y))=
(\delta_l^i\delta_k^j-\delta_k^i\delta_l^j)y_8^2+\frac{1}{2}\ast
Z^{ijab}Z_{lkab} \label{kovar1} \eeq \noindent
 \beq
(G(y),\hat{p}^i\hat{n}_j\hat{p}^k\hat{n}_lG(y))=\delta_l^i\delta_j^k\sum_{n=1}^8y_n
^2-(G(y),\hat{p}^{ik}\hat{n}_{jl}G(y)).
\label{kovar3} \eeq \noindent

After using the results of  Section 5.3. and implementing self
duality for the matrix elements of the basic covariant
${\mathcal{R}^{IJ}}_{KL}$ we get \beq
{\mathcal{R}^{ij}}_{kl}=\left(\delta_l^i\delta_k^j-\delta_k^i\delta_l^j\right)y_8^2
+\sum_{a<b}Z^{ijab}Z_{ablk} \label{Rijkl} \eeq \noindent \beq
{\mathcal{R}^{k+8l+8}}_{i+8j+8}={\mathcal{R}^{ij}}_{kl}
\label{Rplus8} \eeq \noindent \beq
 {\mathcal{R}^{ij}}_{k+8l+8}=
{\mathcal{R}^{i+8j+8}}_{kl}=2y_8Z_{ijkl} \label{rendesplus8} \eeq
\noindent \beq {\mathcal{R}^{ij+8}}_{k+8l}=
\left(\delta_l^i\delta_j^k-\frac{1}{2}\delta_j^i\delta_l^k\right)\sum_{n=1}^8y_n^2
-{\mathcal{R}^{ik}}_{jl}. \label{kevertplus8} \eeq \noindent

\subsection{Polynomial invariants for the generalized SLOCC group}

Using the matrix elements in  Eq.(\ref{eztitteleg}) one can
calculate the invariants $\mathcal{I}_{2p}$. For the special case
of the state of Eq.(\ref{Klasszstate}) these will be polynomials
in the complex amplitudes $y_j, j=1,\dots 8$. As an algebraically
independent set of these polynomials that are invariant under the
nontrivial part of the generalized SLOCC group $G_0={\rm
Spin}(16,\mathbb{C})$ we would like to propose the one  \beq \{
\mathcal{I}_2,\mathcal{I}_8,\mathcal{I}_{12},\mathcal{I}_{14},\mathcal{I}_{18},\mathcal{I}_{20},
\mathcal{I}_{24}, \mathcal{I}_{30} \}. \label{16algindep} \eeq
Note that the order of the algebraically independent polynomials
has been known for a long
time\cite{Antonyan2,Berdjis,Freud,Vinberg}. Indeed carrying out
the calculations a computer check shows that the set of
Eq.(\ref{16algindep}) polynomials is algebraically independent.

In order to motivate our choice and also understand the meaning  of these polynomials
let us first consider another, $240$ element set of elementary
polynomials $e_s(x_1,\dots ,x_8), s=1,2,\dots 240$,  of the form
 \beq \pm x_i\pm x_j,\qquad  \frac{1}{2}(x_i\pm x_2\pm x_3\pm x_4\pm x_5\pm x_6\pm x_7\pm x_8)
\label{e8polinomok}\eeq \noindent where in the second set only an
{\it even} number of minus signs are allowed. In accord with the
(\ref{rom3}) decomposition this $240=112+128$ split of polynomials
corresponds to the root system of the group $E_8$. Let us now
define the polynomials \beq \Pi_{2p}(x)
=\sum_s^{240}[e_s(x)]^{2p},\qquad
2p=2,8,12,14,18,20,24,30\label{Pipol}. \eeq \noindent They form an
alternative set to our polynomials $\mathcal{I}_{2p}(y)$ coming
from the set of Eq.(\ref{16algindep}). A computer check shows that
they are algebraically independent as well.

Observe now that the set of vectors defined by (\ref{Ek1})-(\ref{Ek2}), (\ref{kilogo}) and (\ref{pe8})
defines a Cartan subspace
$\mathfrak{c}$ (i.e. a maximal commutative subspace) of  $\mathfrak{m}\equiv\mathfrak{m}_{128}$ of Eq.(\ref{rom3}).
(For the explicit form of the commutators of the $\mathfrak{e}_8$ Lie-algebra based on the decomposition of (\ref{rom3}) see the paper of Antonyan and Elashvili\cite{Antonyan2}.)
Let $W\equiv W(\mathfrak{c},\mathfrak{e}_8)$ be the Weyl group of $\mathfrak{e}_8$ regarded as a graded algebra.
Then it is known\cite{Vinberg} that the restriction of polynomial functions $\mathbb{C}[\mathfrak{m}]\to \mathbb{C}[\mathfrak{c}]$ induces an isomorphism
$\mathbb{C}[\mathfrak{m}]^{G_0}\to\mathbb{C}[\mathfrak{c}]^W$.
The upshot of these considerations is that if we restrict the (\ref{16algindep}) generating set taken from the space $\mathbb{C}[\mathfrak{m}]^{G_0}$
 of generalized SLOCC invariant polynomials to the generic class represented by our state of Eq.(\ref{Klasszstate})
one obtains some combinations of an algebraically independent set
taken form the space $\mathbb{C}[\mathfrak{c}]^W$ of polynomials
that are invariant under the action of the Weyl group of $E_8$.
Now it is known\cite{Lee} (for alternative choices see
\cite{Mehta,Talamini}) that as an algebraically independent set of
$\mathbb{C}[\mathfrak{c}]^W$ one can take our new polynomials of
Eq.(\ref{Pipol}) constructed from the roots of $\mathfrak{e}_8$.
In order to find the relationship between the Weyl invariant
polynomials of Eq.(\ref{Pipol}) and our set of
Eq.(\ref{16algindep}) restricted to the eight parameter family of
(\ref{Klasszstate}) we have to relate the complex variables $y_j$
and $x_j$ $j=1,\dots 8$. We choose \beq
y_1=\frac{1}{2}(x_1+x_2+x_3+x_4-x_5-x_6-x_7-x_8),\quad
y_2=\frac{1}{2}(x_1+x_2-x_3-x_4-x_5-x_6+x_7+x_8) \label{rconn12}
\eeq \noindent \beq
y_3=\frac{1}{2}(x_1-x_2+x_3-x_4-x_5+x_6-x_7+x_8),\quad
y_4=\frac{1}{2}(x_1-x_2-x_3+x_4-x_5+x_6+x_7-x_8) \label{rconn34}
\eeq \noindent \beq
y_5=\frac{1}{2}(x_1-x_2-x_3+x_4+x_5-x_6-x_7+x_8),\quad
y_6=\frac{1}{2}(x_1-x_2+x_3-x_4+x_5-x_6+x_7-x_8) \label{rconn56}
\eeq \noindent \beq
y_7=\frac{1}{2}(x_1+x_2-x_3-x_4+x_5+x_6-x_7-x_8),\quad
y_8=\frac{1}{2}(x_1+x_2+x_3+x_4+x_5+x_6+x_7+x_8). \label{rconn78}
\eeq \noindent Now let us solve this system of equations to obtain
$x(y)$. Our aim is to find a relation between
$\mathcal{I}_{2p}(y)$ and $\Pi_{2p}(x(y))$ for
$2p=2,8,12,14,18,20,24,30$. Using the results obtained in
Eqs.(\ref{Rijkl})-(\ref{kevertplus8}) a computer calculation shows
that \beq \mathcal{I}_{2p}(y)=(-1)^p2^{2p-1}\Pi_{2p}(x(y)).
\label{naezittjo} \eeq \noindent

Let $\mathfrak{c}$ be the Cartan subspace of $\mathcal{F}_+$. Then
$G_0\mathfrak{c}$ contains an open subset of $\mathcal{F}_+$ and
is dense. From this it follows that any $G_0$ invariant polynomial
on $\mathcal{F}_+$ is determined by its restriction to
$\mathfrak{c}$. Hence our set of Eq.(\ref{16algindep}) can really
be regarded as an algebraically independent set of $G_0$ invariant
homogeneous polynomials. Clearly the extension of these
polynomials to $\mathcal{F}_+$ can be given. The magnitudes of the
polynomials showing up in this set we would like to propose for
the characterization of the entanglement properties of systems of
fermions with eight modes.

A comment on the structure of these invariants is in order.
One can take for example the more general $71$ parameter subset of states as given by Eq.(\ref{alternatesemisimple}).
(Now we refrain from applying
the restrictions of Eqs.(\ref{z12})-(\ref{z567}).)
Then using the explicit form of the matrix elements given by Eqs.(\ref{kovar1})-(\ref{kovar3}) our invariants can explicitly be calculated.
They can be expressed as polynomials in $y_8$ with expansion coefficients given by traces of powers of the
$28\times 28$ matrices $Z$ and $\ast{Z}$. These expressions are even simpler for the $36$ parameter family
of self dual states $\ast Z=Z$.
Notice, however that for the general case featuring all $128$ amplitudes even for the simplest nontrivial invariant, i.e. the octic one $\mathcal{I}_{8}$, one would obtain a rather complicated formula.
We mention that $\mathcal{I}_8$ is living inside the octic $E_8$ invariant calculated in the paper of M. Cederwall\cite{Ceder}.

\section{Conclusions}

In this paper following the ideas of Ref.\cite{SarLev1} we have
been considering the problem of embedding qubits into fermionic
Fock space based on an underlying Hilbert space of dimension $N$.
Unlike previous studies making use of a subspace representing
systems with the number of fermions fixed here we also allowed the
possibility of creating and annihilating fermions via changing
their total number. For this construction to make sense we made
use of the full Fock space. Mathematically this corresponded to
representing pure states of our quantum systems by spinors. This
construction has naturally leads to the idea of extending the
fermionic SLOCC group $GL(N,\mathbb{C})$ to the one of
$G=\mathbb{C}^{\times}\times{\rm Spin}(2N,\mathbb{C})$. In this
picture classification of entanglement types boils down to the
classification of spinors, i.e. the determinations of orbits under
$G$ finding their representatives and their stabilizers. As
emphasized in Ref.\cite{SarLev1} separable states in this
formalism are represented by pure spinors a notion that dates back
to Cartan and Chevalley. Hence entanglement in our new formalism
corresponds to some sort of deviation from purity of spinors.
Though our spinors serving as entangled states are inherently
complex for obtaining real states one can also consider certain
reality conditions. We have shown that a natural reality condition
to be imposed on complex spinors is the one defining Majorana
spinors. The naturality of this condition stems from the fact that
for embedded qubits this condition boils down\cite{SarLevlegujabb}
to the one of self conjugate states under the Wootters spin flip
operation\cite{Wootters}. This operation is of utmost importance
for defining physically well established measures such as the
entanglement of formation for two qubits\cite{Wootters}, and is a
standard ingredient for defining multiqubit measures of
entanglement. It is amusing to see this operation coming out
easily from our Fock space considerations.

Looking at the phenomenon of pure state multipartite entanglement
from our point of view is rewarding from many respects. Here we
elucidated the usefulness of our approach by concentrating on
special entangled systems made of few qubits embedded into Fock
space. We clarified the structure of different types of embedding
via applying the notions of single, double and mixed occupancy.
These notions have transparent physical meaning. We have shown
that the different types of embedding help us to clarify the
physical meaning of structures showing up in the BHQC. The main
problem there was the occurrence of direct sums combined with
tensor products, or the occurrence of singlets apart from
doublets. Though doublets (qubits) have a natural physical
interpretation singlets have no clear cut interpretation within a
conventional framework of entanglement theory. Embedding
entanglement theory to the theory of spinors enables a natural
physical interpretation of singlets.

It is important to note however that we are not pretending that
our ideas solve the problem of singlets showing up in all contexts
featuring the BHQC. Let us consider for instance the problem of
the {\it tripartite entanglement of seven qubits} based on the
$56$ dimensional fundamental irreducible representation of the exceptional group $E_7$
of Refs.\cite{DuffFerrara,Levayfano} related to the work of
Manivel\cite{Manivel}. Since $8\times 7=56$ there the authors
constructed this representation space as the seven-fold direct sum
of $8$ dimensional three-qubit spaces namely \beq V_{ABC}\oplus
V_{ADE}\oplus V_{AFG}\oplus V_{BDF}\oplus V_{BEG}\oplus
V_{CDG}\oplus V_{CEF}.\label{DFLM} \eeq \noindent However, again
without giving a physically sound recipe for what the seven {\it
superselection sectors} in this case mean, this system is left in
a state which is lacking any quantum information theoretic
meaning. Although as demonstrated in Section 5.3. this system
contains seven sectors of $32$ dimensional subspaces amenable to a
femionic interpretation based on the tripartite entanglement of
six qubits the full $56$ dimensional representation space cannot
be embedded into fermionic Fock space. In order to see this just
recall the decomposition of the $56$ of $E_7(\mathbb{C})$ under
$SL(2,\mathbb{C})\times SO(12,\mathbb{C})$ \beq 56=(2,12)\oplus
(1,32). \label{56decomp} \eeq\noindent Here the extra
$SL(2,\mathbb{C})$ factor corresponds to the seventh qubit (say
qubit $G$). The second part $(1,32)$ of this decomposition is
featuring the $32$ dimensional spinor representation amenable to a Fock space
reinterpretation. Its meaning is clearly related to the tripartite
entanglement of six qubits (say $A,B,C,D,E,F$) a picture coming
from the embedded qubits\footnote{There a different labelling of
qubits was used. However, the incidence structure of the
decomposition is the same.} of Section 5.3. However, the first
term is featuring the {\it vector} representation of
$SO(12,\mathbb{C})$ for which no spinorial characterization is
possible. Hence in order to make sense of these constructs from an
entanglement point of view other ideas are needed.

As another application of our ideas we conducted a study on
$n$-qubit invariants reinterpreted as spinorial structures. We
have seen that it is rewarding to enlarge the $n$-qubit SLOCC
group $GL(2,\mathbb{C})^{\times n}$ to $S_n\ltimes
GL(2,{\mathbb{C}})^{\times n}$ by also taking into consideration
permutations of qubits. Being the largest subgroup of the
fermionic SLOCC group $GL(2n,\mathbb{C})$ which leaves invariant
the $n$-qubit subspace spanned by the basis vectors of single
occupancy it serves as a natural group directly related to the
chain \beq S_n\ltimes GL(2,{\mathbb{C}})^{\times n}\subset
GL(2n,\mathbb{C})\subset \mathbb{C}^{\times}\times {\rm Spin}
(4n,\mathbb{C}). \label{chain}\eeq\noindent The rightmost member
of this chain is our generalized SLOCC group of
Eq.(\ref{genslocc}) taken for the special case of $N=2n$. An
important corollary of this observation is that $n$-qubit
invariants featuring also permutation symmetry should be regarded
as ones coming from the basic spinorial invariants and covariants.
In other words for investigating these invariants we should
consider a suitable restriction of the algebra $\mathcal{A}$ of
complex polynomial functions on either $\mathcal{F}_+$ or
$\mathcal{F}_-$. Indeed from a mathematical point of view the
chain of algebras $\mathcal{C}\subset\mathcal{B}\subset
\mathcal{A}$ answering the chain of groups of Eq.(\ref{chain})
should be regarded as the natural object of study. This idea first
appeared in Ref.\cite{Chen} for the inclusion
$\mathcal{C}\subset\mathcal{B}$. In this paper we proposed to
enlarge this inclusion to also include the algebra $\mathcal{A}$.
As an illustration of these ideas we worked out the $n=4$ case.
Here we had the chance to compare our findings with numerous
results already existing in the
literature\cite{Luque,Djok,Djok2,Chen,Gour1,Gour2,Katanova,Levfour}.
The $n=4$ case is also highly special revealing an intriguing
relationship to exceptional groups. Indeed considerations of
Refs.\cite{Gour1,Gour2,Chen} have already revealed that the
algebras $\mathcal{C}$ and $\mathcal{B}$ are related to the
structure of exceptional groups $F_4$ and $E_7$. Via the structure
of the algebra $\mathcal{A}$ our considerations managed to add the
largest exceptional group $E_8$ to the list. In particular we
constructed an algebraically independent set of ${\rm
Spin}(16,\mathbb{C})$ invariant polynomials. The magnitudes of
these polynomials can serve as measures of entanglement in our
fermionic Fock space context. With an explicit computation we have
shown that when restricting these polynomials to the dense orbit
the resulting polynomials on $8$ variables are invariant ones
under $W(E_8)$ i.e. the Weyl group of $E_8$.

\section{Acknowledgements}
One of us (P.L.) would like to express his gratitude to the warm
hospitality at the UTBM in September 2014. This work was supported by a one month visiting
professor grant awarded by the UTBM. We would also like to thank
Professor Elashvili for providing a copy of Ref.\cite{Antonyan2}.
A useful correspondence with V. Talamini is greatly appreciated.

\end{document}